\documentclass[3p, 12pt]{elsarticle}
\usepackage{graphicx}
\usepackage{amsmath}
\usepackage{amsfonts,amssymb}
\usepackage{bm}
\usepackage{psfrag}

\newcommand{\ent}{\eta}                       % entropy
\newcommand{\ie}{e}                           % internal energy
\newcommand{\gfe}{\hat\psi}                   % Gibbs free energy
\newcommand{\hfe}{\psi}                       % Helmholtz free energy
\newcommand{\op}{\varphi}                     % order parameter
\newcommand{\s}{\sigma}                       % stress (scalar)
\newcommand{\sss}{\bm\sigma}                  % stress tensor
\newcommand{\e}{\varepsilon}                  % strain (scalar)
\newcommand{\ee}{\bm\varepsilon}              % strain tensor
\newcommand{\sign}{\textit{sign}}             % sign function
\newcommand{\brac}[1]{\left( {#1} \right)}    % regular brackets (.)
\newcommand{\bracs}[1]{\left[ {#1} \right]}   % square brackets [.]
\newcommand{\bracc}[1]{\left\{ {#1} \right\}} % curly brackets {.}
\newcommand{\vect}[1]{\boldsymbol{#1}}        % vector

\journal{Acta Materialia}

\begin{document}
\begin{frontmatter}
\title{A macroscale, phase-field model for shape memory alloys with non-isothermal
        effects: influence of strain-rate and environmental conditions on the
        mechanical response}
\author[dg]{D. Grandi}
\ead{grandi@dm.unibo.it}
\address[dg]{Department of Mathematics, University of Bologna,
              Piazza di Porta S. Donato 5, I-40126, Bologna, Italy}
\author[mm]{M. Maraldi\corref{cor1}}
\ead{mirko.maraldi@unibo.it}
\address[mm]{DIEM, University of Bologna
              Viale Risorgimento 2, I-40136, Bologna, Italy}
\author[lm]{L. Molari}
\ead{luisa.molari@unibo.it}
\address[lm]{DICAM, University of Bologna
              Viale Risorgimento 2, I-40136, Bologna, Italy}
\cortext[cor1]{Corresponding author}
\begin{abstract}
A Ginzburg-Landau model for the macroscopic behaviour of a shape memory alloy is
proposed. The model is one-dimensional in essence, in that we consider the effect
of the martensitic phase transition in terms of a uniaxial deformation along a
fixed direction and we use a scalar order parameter whose equilibrium values
describe the austenitic phase and the two martensitic variants.
The model relies on a Ginzburg-Landau free energy defined as a function of
macroscopically measurable quantities, and accounts for thermal effects; couplings
between the various relevant physical aspects are established according to
thermodynamic consistency.
The theoretical model has been implemented within a finite-element framework and
a number of numerical tests are presented which investigate the mechanical behaviour
of the model under different conditions; the results obtained are analysed in
relation to experimental evidences available in literature. In particular, the
influence of the strain-rate and of the ambient conditions on the response of
the model is highlighted.
\end{abstract}
\end{frontmatter}

%%%%%%%%%%%%%%%%%%%%%%%%%%%%%%%%%%%%%%%%%%%%%%%%%%%%%%%%%%%%%%%%%%%%%%%%%%%%%%%%
\section{Introduction}
Shape memory alloys [SMA] are materials having many applications and attracting
a lot of interest due to their unique properties of shape memory and
pseudoelasticity, which stem from both temperature-induced and stress-induced
martensitic phase transition. The mechanical behaviour of such materials is
rather complex and arises from a strong interaction between thermal and mechanical
phenomena. In fact, when these materials undergo the martensitic phase transition,
the increase in temperature due to localised self-heating/self-cooling has been
experimentally found to be anything but negligible \cite{leo:1993,shaw:1995,
zhang:2010,he:2010_2}. Hence, the effects of heat transfer and of the heat
dissipation towards the ambient are of primary relevance in the study of these
materials \cite{shaw:1995,he:2010_1} and play a role in the rate-dependent behaviour
of SMA \cite{zhang:2010, tobushi:1998}.

In modelling the mechanical behaviour of a SMA it is therefore important to
account for mechanical and thermal aspects at the same time and to evaluate all
the phenomena in their time evolution.

In literature there exist many constitutive models for SMA \cite{paiva:2006}, which
are derived following different approaches. Some authors propose models developed
within plasticity frameworks, which make use of one or even more internal variables
to describe the pseudo-elastic behaviour \cite{leclerq:1995,lubliner:1996,
auricchio:1997}.
Another group of constitutive models has been developed within a thermomechanic
context; in his work, Tanaka \cite{tanaka:1986} proposed a macroscale model
comprising an internal variable  to quantify the extent of the phase transition,
a dissipation potential and an assumed transformation kinetic.
Similarly, other models based on the same thermodynamic background but featuring
different kinetic laws have been formulated \cite{liang:1997,brinson:1993}, and
a complete heat equation which accounts for the contribution of latent heat and
of the dissipation connected with the phase transition was added to the framework
\cite{brinson:1996,zhu:2007}. Among others, Chang et al. \cite{chang:2006} proposed
a model comprising a strain-gradient elastic free energy and a chemical
(transitional) free energy; Abeyaratne et al. \cite{abeyaratne:1994} developed a
model including a non-convex Helmholtz free energy function of the strain in order
to describe the regions of stability of the different phases.
In the context of using a non-convex free energy, it is not uncommon that
micro-mechanical models are proposed \cite{goo:1997,muller:2001}; Muller and
Seelecke \cite{muller:2001} developed a microscale model based on statistical
physics comprising a bulk free energy function of the lattice shear deformation.

Another approach to the description of the behaviour of SMA consists in applying
the Ginzburg-Landau theory for phase transitions. At a microscopic scale,
Falk \cite{falk:1980} developed a Landau theory based on a shear strain
order parameter. Levitas and Preston \cite{levitas:2002_1} proposed a single-grain
model based on the decomposition of the strain in an elastic and a transformational
part, the latter being described by a pure order parameter which can assume
different values to identify the different phases. In a similar fashion, Wang and
Khachaturyan \cite{wang:1997} proposed a model in which the strain field depends
on several order parameters, whose evolution is described by several time-dependent
Ginzburg-Landau [TDGL] equations. Nevertheless, attempts to handle the problem at
a larger scale of observation have been made as well. Among others, Ahluwalia et
al. \cite{ahluwalia:2003} and Chen and Yang \cite{chen:1994} proposed meso-scale
models for the description of polycrystalline materials in which the order parameters
account for the different orientations of each single grain; in the same spirit,
Brocca et al. \cite{brocca:2002} proposed a microplane model attempting to bridge
the gap between micromechanics-based and macroscale models.
Berti et al. \cite{berti:2010_2} proposed a Ginzburg-Landau model which can be
applied at the macroscale and encompasses mechanical as well as thermal effects by
introducing the balance of linear momentum equation and the heat equation in a
thermodynamically consistent framework; the order parameter is related to the extent
of the phase transition between austenite and the martensite variants and its
evolution is regulated by a TDGL equation. Owing to the presence of the heat equation,
it is possible to describe the thermo-mechanical interactions which strongly
influence the constitutive behaviour of a SMA and to account for non-isothermal
conditions. The model is presented both in a three-dimensional and a monodimensional
setting; the reduction to this latter case is still meaningful because in most
engineering applications SMA wires are employed.

In this paper, we present a numerical study on the thermomechanical behaviour of
a SMA, with a particular focus on the rate-dependent response and on the influence
of thermal conduction and heat transfer in the mechanical behaviour. Starting
from the models proposed in Berti et al. \cite{berti:2010_2} and Daghia et al.
\cite{daghia:2010}, we formulate a new free energy functional and give a proper
expression for the relaxation parameter regulating the TDGL equation; this gives
the model the properties required to reproduce the symmetrical behaviour between
the austenite-to-martensite and the martensite-to-austenite phase transitions
observed experimentally.
The ability of the model to quantitatively reproduce a variety of experimental
evidences of a typical polycrystalline NiTi alloy is then demonstrated.

The paper is organised as follows. In Section \ref{sect:model}, the theoretical
model is described, starting from a suitable Ginzburg-Landau free energy whose
properties are outlined in Section \ref{sect:GL-pot}. Constitutive relations are
given in Section \ref{sect:constit}, while the thermodynamic consistency of the
model is shown in Section \ref{sect:consistency} and the complete differential
system with appropriate boundary and initial conditions is provided in Section
\ref{sect:equations}.
In Section \ref{sect:galerkin} the Galerkin formulation of the differential
problem, suitable for the finite element implementation, is sketched.
Section \ref{sect:num-tests} is the main part of the paper, where the numerical
results of a number of simulated tensile tests on a bar specimen under
different conditions are reported. After describing, in Section \ref{sect:baseline},
the effectiveness of the model in recovering the main experimental evidences in
terms of stress-strain response, phase morphology and temperature evolution, in
Section \ref{sect:temp-test} tensile tests performed at different values of the
initial temperature are illustrated, while in Section \ref{sect:max-def} the
stress-strain response under partial-loading conditions is depicted. The rate-dependent
behaviour of the model is investigated in Section \ref{sect:strain-rate}, with a
particular focus on the effects of the strain-rate on the domain nucleation, the
hysteresis cycle and the energy dissipation. The influence of heat transfer phenomena
on the mechanical response of the specimen is examined in Section \ref{sect:environment}.
The last two aspects are considered jointly in Section \ref{sect:env-rate}.
The paper ends drawing some conclusions in Section \ref{sect:concl}.

%%%%%%%%%%%%%%%%%%%%%%%%%%%%%%%%%%%%%%%%%%%%%%%%%%%%%%%%%%%%%%%%%%%%%%%%%%%%%%%%
\section{Model}
\label{sect:model}
The thermomechanical properties of SMA stem from the nature of the martensitic
phase transition; here we will account for it by means of a Ginzburg-Landau
approach \cite{falk:1980,levitas:2002_1}. To this end we have to define a
phase field, or order parameter, $\op$; differently from the approach followed,
among others, by Falk \cite{falk:1980}, the order parameter is not identified
with the uniaxial strain $\e_{11}$, but rather the phase field $\op$ is used here
as a macroscopic indicator of the phase (martensite or austenite) of the material
at every point \cite{levitas:2002_1, berti:2010_2, artemev:2000, artemev:2001, berti:2010_1}.
Strain is an independent field, though it is coupled with the order parameter.
We assume that the martensite phase be present in only two relevant variants,
$M_\pm$, characterised by opposite uniaxial transformation strains with respect
to the austenite phase [A], namely $\pm\bm\varepsilon_t = \pm\e_0 \bm n \otimes \bm n$,
where $\bm n$ is a fixed direction. The Ginzburg-Landau potential, whose properties
will be described in detail in Section \ref{sect:GL-pot}, is expressly formulated to
have (at most) three stable values of $\op$ at values $\bracc{0,\pm1}$, corresponding
to phases $A,M_\pm$, respectively. We remark that, as a consequence of the
\emph{first order} character of the martensitic phase transition, intermediate
values of $\op$ do not represent any physically stable phase, but only transition
layers between different stable or metastable physical phases.

The model has to encompass evolution equations for the three basic fields: the
order parameter ($\op$), the strain ($\bm\varepsilon$) and the temperature ($\theta$).
The evolution of the phase field will be described by the TDGL equation
\begin{equation}\label{eqn:GL-eq}
 \tau(\op)\dot\op=-\frac{\partial\gfe}{\partial\op}+\nabla\cdot\left(\frac{\partial\gfe}{\partial(\nabla\op)}\right)
\end{equation}

The strain will be determined through the combination of the balance of linear
momentum equation and a proper constitutive relation between strain, stress and
order parameter, while the heat equation regulating the temperature evolution will
arise from the balance of energy equation.

For the formulation of the model, we consider a bar sample of SMA occupying
a bidimensional material domain $\Omega$. We indicate with $x=(x_1,x_2)$ the
material coordinates and with $\mathbf u(x)$ the displacement vector field.
Relying on the small displacement approximation, we use the linearised strain tensor
$\bm\varepsilon = \nabla^S \mathbf u = \frac{1}{2}[\nabla\mathbf u+(\nabla\mathbf u)^T]$.

%-------------------------------------------------------------------------------
\subsection{Ginzburg-Landau potential}
\label{sect:GL-pot}
Following Berti et al. \cite{berti:2010_2}, we assume a Ginzburg-Landau free
energy in the form
\begin{eqnarray} \label{Gibbs1D}
\gfe(\theta, \sss, \op, \nabla \op) &=& - \frac{c}{2 \theta_{ref}} \theta^2
                                        - \frac{\lambda}{2} \sss : \sss
                                        + \frac{\kappa}{2} |\nabla\op|^2  + \nonumber\\
                                     && + \frac{\ell}{2} \bracs{\brac{\theta_A-\theta_M} F(\op)
                                                        + \brac{\hat\theta - \frac{\bm\varepsilon_t :\sss}{\ell} \sign(\op)} G(\op)},
\end{eqnarray}
where the temperature function $\hat\theta$ is given by
\begin{equation} \label{theta-hat}
 \hat\theta = \max \bracc{\brac{\theta - \theta_A}, \brac{\theta_0 - \theta_A}},
                        \qquad \theta_0 < \theta_M < \theta_A,
\end{equation}
being $c$ the specific heat, $\theta_{ref}$ a reference temperature for the specific
heat, $\lambda$ the compliance modulus, $\sss$ the stress tensor, $\kappa$ the
interface energy parameter regulating the interface width, and $\ell$, $\theta_A$,
$\theta_M$, $\theta_0$ material parameters related to the phase transition whose
physical meaning will be explained in the following.

The two potentials $F$ and $G$ are defined by
\begin{eqnarray}
  F(\op) &=& \frac{1}{2} \bracs{1 - \cos \brac{\pi \op}} + \frac{1}{8} \bracs{1 - \cos \brac{2 \pi \op}},\\
  G(\op) &=& 1 - \cos \brac{\pi \op}.
\end{eqnarray}
\begin{figure} \begin{center}
\begin{tabular}{cc}
  \includegraphics[scale=0.25]{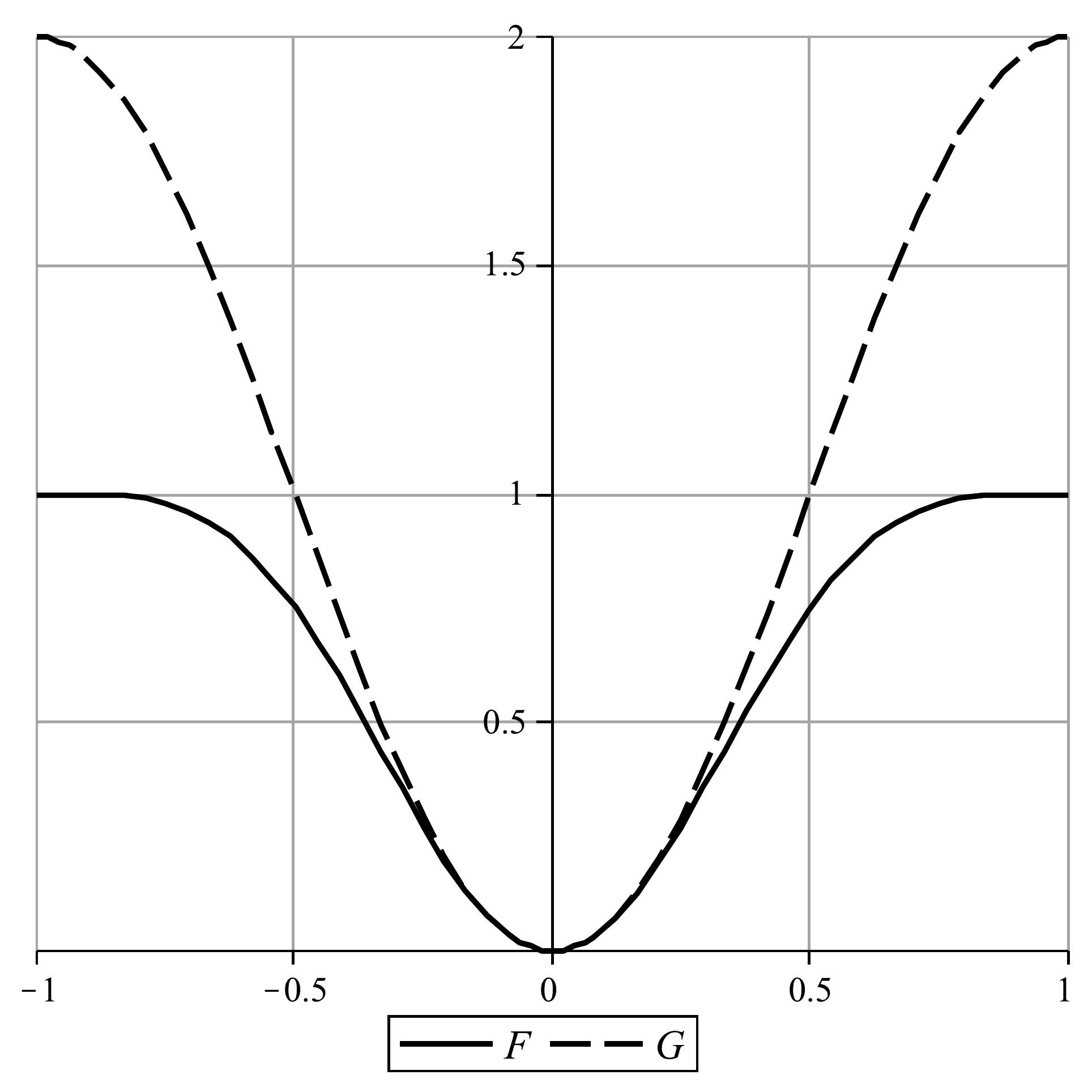} & \includegraphics[scale=0.25]{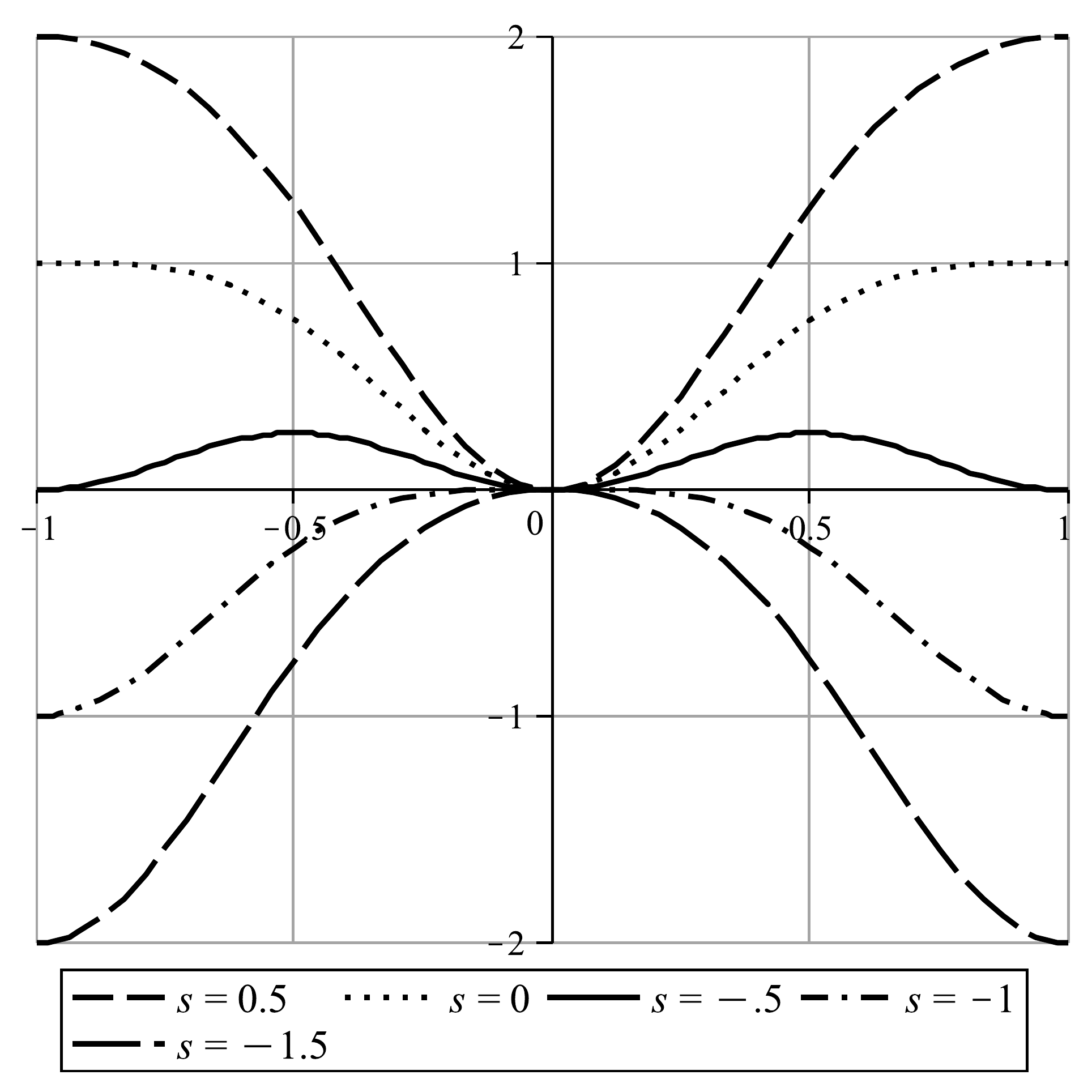} \\
  (a) & (b)
\end{tabular}
\caption{(a) Functions $F(\op)$ and $G(\op)$ defining the free energy. (b) The potential
  $W(\op; s)$ in equation \eqref{eqn:W} for different values of the parameter $s$.}
\label{fig:potentials}
\end{center} \end{figure}
and are depicted in Fig. \ref{fig:potentials}(a). The picture shows the shape of
the potentials in the interval $[-1,1]$, which is the physically possible range
of the order parameter.

These potentials are indeed different from the polynomial ones usually employed
in Landau theory, in particular in Berti et al. \cite{berti:2010_2,berti:2010_1}.
The choice we made here was suggested by a preliminary numerical investigation,
which showed the drawbacks of using polynomial potentials. To clarify this issue,
we first point out that the functions $F$ and $G$ are defined in a way that, in
the interval $\bracs{-1, 1}$, the function
\begin{equation} \label{eqn:W}
 W(\op; s) = F(\op) + s \ G(\op)
\end{equation}
has minima only at $\op = 0$ or at $\op = \pm1$ for any value of the parameter
$s \in \mathbb{R}$. More precisely if $s \geq 0$, $W$ has an (absolute) minimum
in $\op = 0$ and two inflection points in $\op = \pm1$. For $s \leq -1$ there are
two absolute minima at $\op = \pm1$ and a relative maximum at $\op = 0$. If
$-1 < s < 0$ there are three relative minima located at $\op = 0, \pm1$; in
particular, for $s = - 1/2$, the three minimal points have equal values. Similar
properties can be achieved using sixth order polynomials. However, by using
trigonometric functions it is possible to obtain a more symmetrical potential as
it regards to the properties of the minima in the three points $\op = 0,\pm1$.
More precisely, using trigonometric potentials, the following property holds
\begin{equation} \label{eqn:inversion}
 W(\op \pm1; -s - 1) = W(\op; s) - 1 - 2 s.
\end{equation}
In particular, $W(\op \pm 1; -1) = W(\op; 0) - 1$ (see Fig. \ref{fig:potentials}(b));
this means that, at the critical value $s = - 1^+$, the shape of the  incipient
maximum at $\op = 0$ is identical to that of the incipient maxima at $\op = \pm1$ for
$s = 0^-$. As a consequence, the driving force for the phase transitions
$A\rightarrow\pm M$ and $\pm M\rightarrow A$ are the same. This circumstance
allows to obtain a symmetrical behaviour of the stress-strain curves in the loading
stage with respect to the unloading stage which would not be possible to achieve
with polynomial potentials.

\begin{figure}[!h]
\begin{center}
\includegraphics[scale=0.4, trim = 0cm 3cm 0cm 6cm, clip=true]{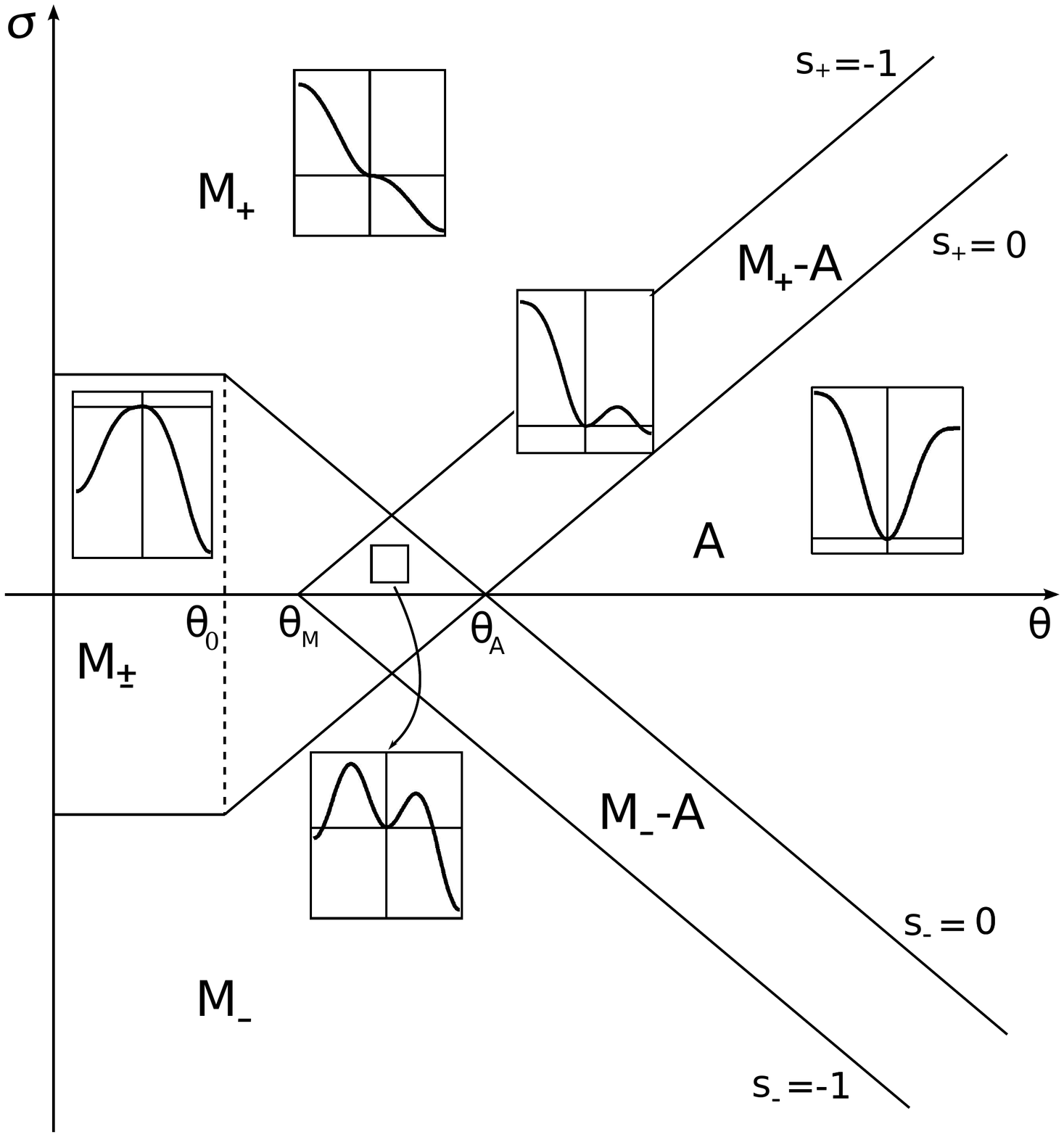}
\end{center}
\caption{The equilibrium phase diagram resulting from local free energy minimisation.
          In the insets the potentials $W(\op; s)$ in the range $-1 \leq \op \leq 1$
          for the values of $s$ corresponding to the selected region of the plane
          $\theta-\sigma$ are depicted.}
\label{fig:phase-d}
\end{figure}

The structure of the minima of the free energy function \eqref{Gibbs1D} in the
half-lines $\op \geq 0$ and $\op \leq 0$ is determined respectively by the parameters
$$
 \hat{s}_{\pm} = \frac{1}{\theta_A-\theta_M} \brac{\hat\theta \mp \frac{\bm\varepsilon_t : \sss}{\ell}}
$$
according to the following rules:
\begin{itemize}
 \item [-] if $\hat s_{\pm} > 0$, then $\op = \pm1$ is a relative maximum (unstable),
 \item [-] if $\hat s_{\pm} < 0$, then $\op = \pm1$ is a relative minimum (stable or metastable),
 \item [-] if $\hat s_+ < - 1$ or $\hat s_- < -1$, then $\op = 0$ is a relative maximum or a inflection point (unstable),
 \item [-] if $\hat s_+ > - 1$ and $\hat s_- > -1$, then $\op = 0$ is a relative minimum (stable or metastable).
\end{itemize}
Thus the equations
$$
 \hat{s}_{\pm}(\theta, \s) = -1, \qquad \hat{s}_{\pm}(\theta, \s) = 0
$$
give rise to piecewise straight lines in the $\theta-\s$ plane (see Fig. \ref{fig:phase-d}).
The temperature $\theta_0$ has to be chosen lower than $\theta_M$, otherwise
$\hat\theta > \theta_M$ at any temperature and the local minimum at $\op = 0$
would never disappear (so the phase $A$ would always be metastable, even at
$\theta < \theta_M$).

The three independent temperatures $\theta_A$, $\theta_M$, $\theta_0$, as well as
the energy $\ell$ and the strain $\e_0$ are adjustable parameters for the
transition. The ratio $\ell / \e_0$, gives the slope of the straight lines in the
$\sigma-\theta$ plane, while $\e_0$ is the transformation strain.

%-------------------------------------------------------------------------------
\subsection{Stress-strain constitutive relation and thermodynamic potentials}
\label{sect:constit}
After having described the Ginzburg Landau potential which appears in the phase
field equation \eqref{eqn:GL-eq}, we give the phase dependent stress-strain
constitutive relation which complements the balance of linear momentum equation.

Following Levitas and Preston \cite{levitas:2002_1} and Berti et al. \cite{berti:2010_2},
we assume the following constitutive relation:
\begin{equation} \label{eqn:stress-strain}
  \bm \varepsilon = - \frac{\partial \hat \psi}{\partial \sss}
                  = \lambda \sss + \frac{\bm\varepsilon_t}{2} \sign(\op) G(\op).
\end{equation}
The tensor $\bm\varepsilon_t = \e_0 \bm n \otimes \bm n$ describes a deformation
in the $\bm n$ direction. From this relation, recalling that $G(1) - G(0) = 2$,
it can be noted that $\bm\varepsilon_t$ represents the strain difference between
the phase $\op=1$ and $\op=0$ at constant stress (\emph{transformation strain}).

In Section \ref{sect:GL-pot} we noticed that $\ell / \e_0$ corresponds to the slope
of the transition lines in the stress-temperature phase space. We will see now
that $\ell$ is also related to the latent heats of the transitions. To this end,
we define the entropy function as
\begin{equation}
 \ent = - \frac{\partial \hat\psi}{\partial\theta}
      =   \frac{c}{\theta_{ref}} \theta
        - \frac{\ell}{2} \frac{\partial\hat\theta}{\partial\theta}G(\op)
      = \frac{c}{\theta_{ref}} \theta - \frac{\ell}{2} H(\theta - \theta_0)G(\op).
\end{equation}
This definition follows from standard thermodynamics arguments, assuming that $\hat\psi$
corresponds to the Gibbs free energy. In Section \ref{sect:consistency} we will show
that this assumption and the subsequent expression of $\ent$ (as well as of the
internal energy $e$) are in agreement with the second law of thermodynamics. The
latent heat for the  transition from an initial phase $i$ to a final one $f$ is
\begin{equation}
 L_{i\rightarrow f} = \theta_{tr} \brac{\ent_f-\ent_i},
\end{equation}
where $\theta_{tr} \geq \theta_M$ is the transition temperature; given
this definition, $L > 0$ means that a positive amount of heat is absorbed by the
system. For $\s = 0$ the transitions occur at temperatures $\theta_A$ and $\theta_M$;
in particular we have
\begin{equation}
  L_{A \rightarrow M \pm} = - \theta_M \ell, \quad
  L_{M \pm \rightarrow A} =   \theta_A \ell .
\end{equation}
As the Ginzburg-Landau potential $\gfe$ corresponds to the Gibbs free energy, the
Helmholtz free energy is defined by
\begin{equation}
 \hfe = \gfe + \s \ee = -\frac{c}{2\theta_{ref}}\theta^2 + \frac{\lambda}{2} \s^2
      + \frac{\kappa}{2} |\nabla\op|^2
      + \frac{\ell}{2} \bracs{\brac{\theta_A - \theta_M} F(\op) + \hat\theta G(\op)},
\end{equation}
We note that $\ent= - \hfe_\theta$ also holds. Finally, the internal energy is
accordingly given by
\begin{equation}
 \ie = \hfe + \theta \ent
     = \frac{c}{2\theta_{ref}}\theta^2 + \frac{\lambda}{2} \s^2 + \frac{\kappa}{2} |\nabla\op|^2
     + \frac{\ell}{2} \bracs{\brac{\theta_A - \theta_M} F(\op) + \brac{\hat\theta - \theta \hat\theta'} G(\op)}.
\end{equation}

%-------------------------------------------------------------------------------
\subsection{Thermodynamic consistency}
\label{sect:consistency}

In this section, we show that the constitutive stress-strain relation and the
expression of the thermodynamic potential are consistent with the second law of
thermodynamics in the form of the Clausius-Duhem inequality
\begin{equation}
\dot \ent \geq - \nabla\cdot\left( \frac{\mathbf q}{\theta}\right) + \frac{r}{\theta},
\end{equation}
where $\mathbf q$ is the heat flux and $r$ the external heat supply. To this end,
an appropriate form of the first law has to be stated:
\begin{equation} \label{eq:bal-nrg}
 \dot e = \mathcal P_{\op}^i + \mathcal P_m^i - \nabla \cdot \mathbf q + r,
\end{equation}
where $\mathcal P_m^i$ and $P_{\op}^i$  are the internal powers associated,
respectively, to the balance of linear momentum equation
\begin{equation}
 \rho \ddot {\mathbf u} = \nabla \cdot \bm\sigma + \rho \mathbf b,
\end{equation}
and to the Ginzburg Landau equation \eqref{eqn:GL-eq}. Exploiting the first law
to eliminate the source term in the Clausius-Duhem inequality and recalling the
definition of the Helmholtz free energy $\psi = e - \theta \ent$, the reduced
inequality is obtained
\begin{equation}
\dot \hfe + \ent \dot{\theta} - \mathcal P_{\op}^i - \mathcal P_m^i
     + \frac{\mathbf q}{\theta} \cdot\nabla\theta\leq 0.
\end{equation}
For the internal mechanical power we adopt the standard expression
\begin{equation} \label{eq:mech-power}
 \mathcal P_m^i=\bm\sigma:\dot{\bm\varepsilon}.
\end{equation}
The power balance associated to the Ginzburg-Landau equation is obtained by
multiplying that equation by $\dot\op$ \cite{fried:1993,fabrizio:2006,fabrizio:2008}:
\begin{equation}
 \tau \dot\op^2 = - \gfe_\op \dot\op + \dot\op \nabla \cdot \brac{\kappa \nabla \op}
                = - \gfe_\op \dot\op - \kappa \dot \nabla \op \cdot \nabla \op
                  + \nabla \cdot \brac{\kappa \dot\op \nabla \op}
\end{equation}
The internal power is then defined by
\begin{equation} \label{eq:op-power}
 \mathcal P_\op^i = \tau \dot\op^2 + \gfe_\op \dot\op + \kappa \dot \nabla \op \cdot \nabla \op,
\end{equation}
while $\mathcal P_\op^e = \nabla \cdot \brac{\kappa \dot\op \nabla \op}$ is the balancing
external power.

By substituting the power expressions and the Helmholtz free energy
$\psi = \hat\psi + \sss : \bm\varepsilon$ in the reduced equality, we obtain the
condition
\begin{equation}
    \brac{\hat\psi_\theta + \ent} \dot\theta + \brac{\hat\psi_{\bm\sigma} - \bm\varepsilon} : \dot{\bm\sigma}
  - \tau \dot\op^2 + \frac{\mathbf q}{\theta} \cdot \nabla \theta \leq 0,
\end{equation}
which has to hold for every process. This is satisfied by the given constitutive
stress-strain relation $\bm\varepsilon = \hat\psi_{\bm\sigma}$ and entropy
relation $\ent = - \hat\psi_\theta$, with the further conditions
\begin{equation}
 \tau > 0 \quad \textrm{and} \quad \mathbf q = - k \nabla \theta, \quad k > 0.
\end{equation}
Using the expressions of the internal powers \eqref{eq:mech-power} and \eqref{eq:op-power},
we also write in an explicit way the balance of energy \eqref{eq:bal-nrg}:
\begin{equation}
    \frac{c \ \theta}{\theta_{ref}} \dot\theta
  - \frac{\ell}{2} \bracs{\theta \hat\theta''G(\op) \dot\theta + \theta \hat\theta'\dot G(\op)}
  - \tau \dot \op^2 = k \Delta \theta + r.
\end{equation}

%-------------------------------------------------------------------------------
\subsection{Differential system and boundary conditions}
\label{sect:equations}
In this section we collect the equations of the model and state appropriate boundary
and initial conditions for the problem.

As it regards the balance of linear momentum
equation, we will adopt the quasi-static approximation neglecting the inertial
term $\rho \ddot{\mathbf u}$. This corresponds to assume that the time scales of
thermal phenomena and phase evolution are larger than the stress wave time scale.
We also consider experimental conditions in which the external momentum source
$\mathbf b$ and heat source $r$ vanish. We obtain
\begin{eqnarray} \label{eq:model}
\left\{\begin{array}{ll}
\nabla\cdot\bm\sigma=0,\\
 \tau(\op) \dot{\op} + \frac{\ell}{2}\bracs{\brac{\theta_A-\theta_M} F'(\op)
                             + \brac{\hat\theta - \frac{\e_0}{\ell}  \bm n \otimes \bm n : \sss \sign(\op)} G'(\op)}
               - \kappa \Delta\op = 0,\\
\bracs{c \frac{\theta}{\theta_{ref}} -  \frac{\ell}{2} \theta \ \delta(\theta-\theta_0) G(\op)} \dot\theta
  - \frac{\ell}{2} \ \theta \ H(\theta-\theta_0) G'(\op) \dot\op
  - \tau \dot \op^2 =  k\Delta\theta,
\end{array}\right.
\end{eqnarray}
with
\begin{eqnarray}
  \lambda\bm\sigma = \nabla^S \mathbf u - \frac{\e_0}{2} \bm n \otimes \bm n \sign(\op) G(\op),\\
 \tau(\op) = \tau_{min} +\frac{1}{2} \brac{\tau_{max} - \tau_{min}}[1- \cos(2\pi\op)].
\end{eqnarray}

The relaxation parameter $\tau$ in the Ginzburg-Landau equation has been chosen on
the basis of the experimental data available in literature \cite{leo:1993,
zhang:2010, he:2010_2}.

We now state the boundary and initial conditions accompanying the differential
system. We consider a two-dimensional rectangular domain $\Omega = [0,L] \times [0,a]$.
With $\partial\Omega_+, \partial\Omega_-$ and $\partial\Omega_1, \partial\Omega_2$
we denote two different partitions of the boundary $\partial \Omega$. At any point
of the boundary, $\bm\nu$ denotes the outward unit normal. Typical boundary
conditions are
\begin{eqnarray}
 \nabla\op(x,t)\cdot\bm \nu|_{\partial\Omega} &=& 0, \label{eq:ins-BC}\\
  u(x,t)|_{\partial\Omega_-} &=& \tilde u(x),\\
\nabla u(x,t)\cdot\bm \nu|_{\partial\Omega_+} &=& t(x),\\
 \theta(x,t)|_{\partial\Omega_1} &=& \tilde \theta(x),\\
\nabla\theta(x,t)\cdot\bm \nu|_{\partial\Omega_2} &=& g(x) \label{eq:heat-flux-BC}.
\end{eqnarray}
We remark that equation \eqref{eq:ins-BC} represents the boundary `insulation
condition', which states that no long distance interactions are
allowed between the domain and its exterior \cite{polizzotto:2003}; Dirichlet
boundary conditions are not physically meaningful for the phase parameter.
Mixed Dirichlet and Neumann boundary conditions are instead common for the
temperature and the displacement fields. In the numerical tests, a different form
for equation \eqref{eq:heat-flux-BC} will be used:
$$
k\nabla\theta(x,t)\cdot\bm \nu|_{\partial\Omega_2}=h(\theta-\theta_{ext}),
$$
which is a convective heat transfer condition, representing a situation in which
a boundary heat flux driven by the temperature difference with the external
environment (via the exchange coefficient $h$) is prescribed.

As it regards the initial conditions, they have to be set only for the temperature
and the phase variable:
\begin{equation}
 \theta (x, 0) = \theta_0(x),\quad  \op (x, 0) = \op_0(x),
\end{equation}
as for the displacement field $u$ we are using the quasi-static equation
(\ref{eq:model}$.1$).

%%%%%%%%%%%%%%%%%%%%%%%%%%%%%%%%%%%%%%%%%%%%%%%%%%%%%%%%%%%%%%%%%%%%%%%%%%%%%%%%
\section{Galerkin formulation of the problem}
\label{sect:galerkin}
In this section, a formulation of the governing equations is presented which will
be used in computing numerical examples.

To formulate the finite element problem,
let $\mathcal{T} = \bracc{E}$ be a triangulation of the domain $\Omega$ into finite
element cells. We will work with the usual Lagrange finite element basis with the
definition of the following piecewise polynomial spaces:
\begin{equation}
  W_{k} = \bracc{v^h \in H^{1}\brac{\Omega}, \left.v^h\right|_E \in P_{k}\brac{E} \forall E \in \mathcal{T}}, \notag
\end{equation}
\begin{equation}
  V_{k} = \bracc{ v^h \in W_k : v^h=0\; {\text on}\; \partial\Omega}, \notag
\end{equation}
where $P_{k}$ denotes the space of Lagrange polynomials.  The numerical
formulation that we propose involves the solution of the following problem:
given the data at time $t_{n}$, find $\brac{{\bf u}^{h}, \op^{h}, \theta^{h}}
\in (W_{k})^{2} \times W_{k} \times W_{k}$ (where $k > 0$) at time $t_{n+1}$ such that
\begin{eqnarray}
L_{\mathbf{u}^{h}} & = & \int_{\Omega}{\nabla{\mathbf w^h} \cdot \frac{1}{\lambda}\brac{\nabla \mathbf u_m^h - \frac{\bm\varepsilon_t}{2} \sign(\op_m^h) G(\op_m^h)} dx} =0  \;
                                \;\;\; \forall {\mathbf w^h} \in (V_{k})^{2} \\
L_{\op^{h}}        & = & \int_{\Omega}{q^h \frac{\ell}{2} \bracs{ \brac{\theta_A-\theta_M} F'(\op_m^h)
                                                             + \brac{\hat\theta - \frac{\bm\varepsilon_t : \bm\sigma_m^h}{\ell} \sign(\op_m^h)} G'(\op_m^h)} }dx + \\ \notag
                   & & + \int_{\Omega}{q^h \tau(\op_m^h) \frac{\op_{n+1}^h-\op_{n}^h}{\Delta t} }dx
                       + \int_{\Omega}{\nabla q^h \cdot \kappa (\nabla \op^h)_m}dx = 0
                                \;\;\; \forall q^h \in V_{k} \\
L_{\theta^h}       & = & \int_{\Omega}{z^h \bracs{c \frac{\theta_m^h}{\theta_{ref}} - \frac{\ell}{2} \theta_m^h \ \delta(\theta_m^h-\theta_0) G(\op_m^h)}
                                \frac{\theta_{n+1}^h-\theta_{n}^h}{\Delta t} }dx + \\ \notag
                   & & - \int_{\Omega}{z^h \frac{\ell}{2} \theta_m^h \ H(\theta_m^h-\theta_0) G'(\op_m^h) \frac{\op_{n+1}^h-\op_{n}^h}{\Delta t} }dx
                       + \int_{\Omega}{\nabla z^h \cdot k \brac{\nabla\theta^h}_m }dx + \\ \notag
                   & & - \int_{\Omega}{z^h \tau(\op_m^h) \brac{\frac{\op_{n+1}^h-\op_{n}^h}{\Delta t}}^2 }dx
                       + \int_{\partial\Omega}{z^h h \brac{\theta_m^h-\theta_{ext}} }ds=0
                                \;\;\; \forall z^h \in V_{k},
\end{eqnarray}
where $(\cdot)_m=(1-\beta)(\cdot)_{n+1}+\beta(\cdot)_n$.

$L_{{\bf u}^{h}}$, $L_{\op^{h}}$ and $L_{\theta^h}$ represent the
weak forms of the three governing equations: the balance of linear momentum,
the TDGL equation and heat equation, respectively. The forms are cast, following
a standard process, using the Crank-Nicolson method for the time derivatives;
as the functionals $L_{{\bf u}^{h}}$, $L_{\op^{h}}$ and $L_{\theta^h}$ are nonlinear
in $(\vect{u}^{h}, \op^{h}, \theta^{h})$, a modified Newton method is used to
solve the nonlinear problem at each time step.
In the numerical simulations, linear Lagrange functions are used for the phase and
the temperature fields, while quadratic Lagrange functions are used for the
displacement field.

The computer code used to perform all the simulations has been generated automatically
from a high-level implementation by using a number of tools from the FEniCS Project
\cite{logg:2010}.

%%%%%%%%%%%%%%%%%%%%%%%%%%%%%%%%%%%%%%%%%%%%%%%%%%%%%%%%%%%%%%%%%%%%%%%%%%%%%%%%
\section{Numerical simulations}
\label{sect:num-tests}
A number of numerical simulations were performed with the aim to describe the
effectiveness of the model in capturing the behaviour of a SMA not only from a
qualitative, but also from a quantitative point of view. The results obtained are
presented in this section and their consistency with the experimental evidences
available in literature is highlighted.
\begin{figure}
\begin{center}
  \includegraphics[scale=0.6]{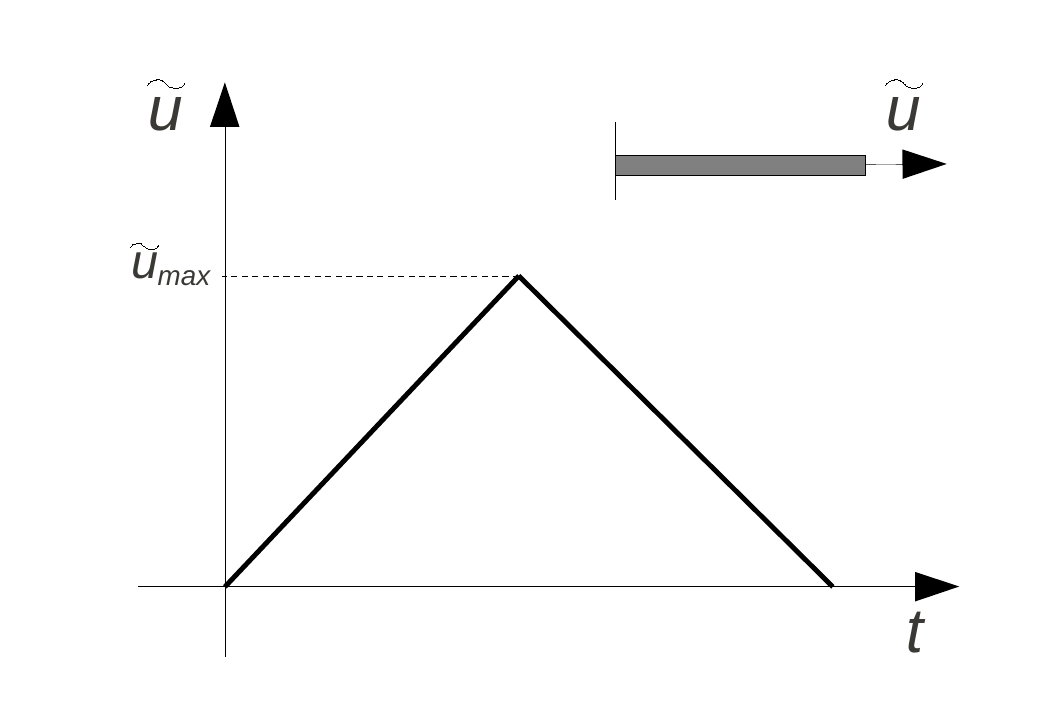}
\end{center}
\caption{Imposed displacement at the end of the bar as a function of time.}
\label{fig:test}
\end{figure}

To reproduce the results of a tensile test conducted on a SMA wire or on a
\textit{dog-bone} specimen, a displacement-controlled uniaxial test was simulated
considering a domain of dimensions $L = 30 \text{ mm}$ and $a = 0.08 \text{ mm}$.
The $\bm n$ direction appearing in the transition tensor $\bm\varepsilon_t$ is
assumed to be along the major domain edge $L$.
The imposed displacement varied linearly from $\tilde{\bm u} = \bm 0$ to a maximum
value $\tilde{\bm u} = \tilde{u}_{max} \bm n$ (loading stage) and then back to
$\tilde{\bm u} = \bm 0$ (unloading stage), as depicted in Fig. \ref{fig:test}.
In all cases, the initial temperature is set equal to the external temperature
($\theta_{ext} = \theta_0$)

To explore the potential of the model in capturing the experimental evidences,
the values of the parameters used in the simulations were chosen according to the
data reported in the paper by He and Sun \cite{he:2010_2} regarding a commercial
polycrystalline NiTi sheet. The parameters for which it was not possible to directly
find a value in the aforementioned paper were chosen in order to reproduce in a
satisfactory way the response observed by Zhang et al. \cite{zhang:2010}, especially
in terms of transition stress, propagation stress, number of nucleating domains and
maximum increase in temperature during the test.

The values of the parameters are: $\lambda = 2.5 \times 10^{-11} \text{ Pa}^{-1}$;
$\varepsilon_0 = 0.058$; $l = 10^{6} \text{ Pa/K}$;  $\theta_A = 288.5 \text{ K}$;
$\theta_M = 273.0 \text{ K}$; $\theta_0 = 212.7 \text{ K}$; $\kappa = 0.15 \text{ N}$;
$\tau_{max} = 8 \times 10^{6} \text{ Pa s}$; $\tau_{min} = 1.6 \times 10^{5} \text{ Pa s}$;
$c = 3.2 \times 10^{6} \text{ Pa/K}$; $\theta_{ref} = 296 \text{ K}$;
$k = 18 \text{ W/(m K)}$.

The values of the heat transfer coefficient $h$, of the initial temperature $\theta_0$
and of the imposed strain rate $\dot{\epsilon}$ varied depending on the test.

%-------------------------------------------------------------------------------
\subsection{General features of a tensile test}
\label{sect:baseline}
\begin{figure} [h!]
\begin{center}
\begin{tabular}{cc}
  (a)&\includegraphics[scale=0.5]{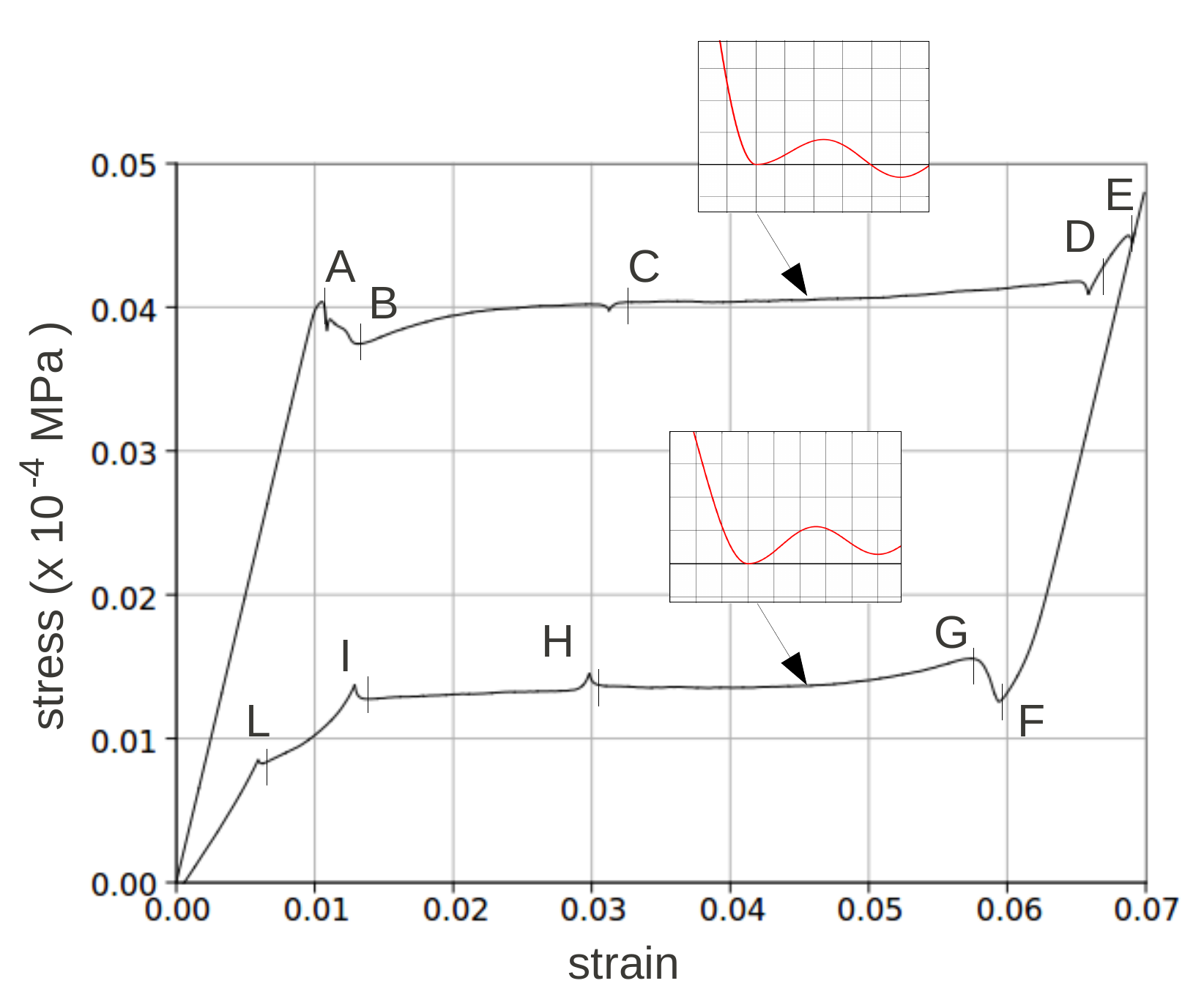} \\
  (b)&\includegraphics[scale=0.4]{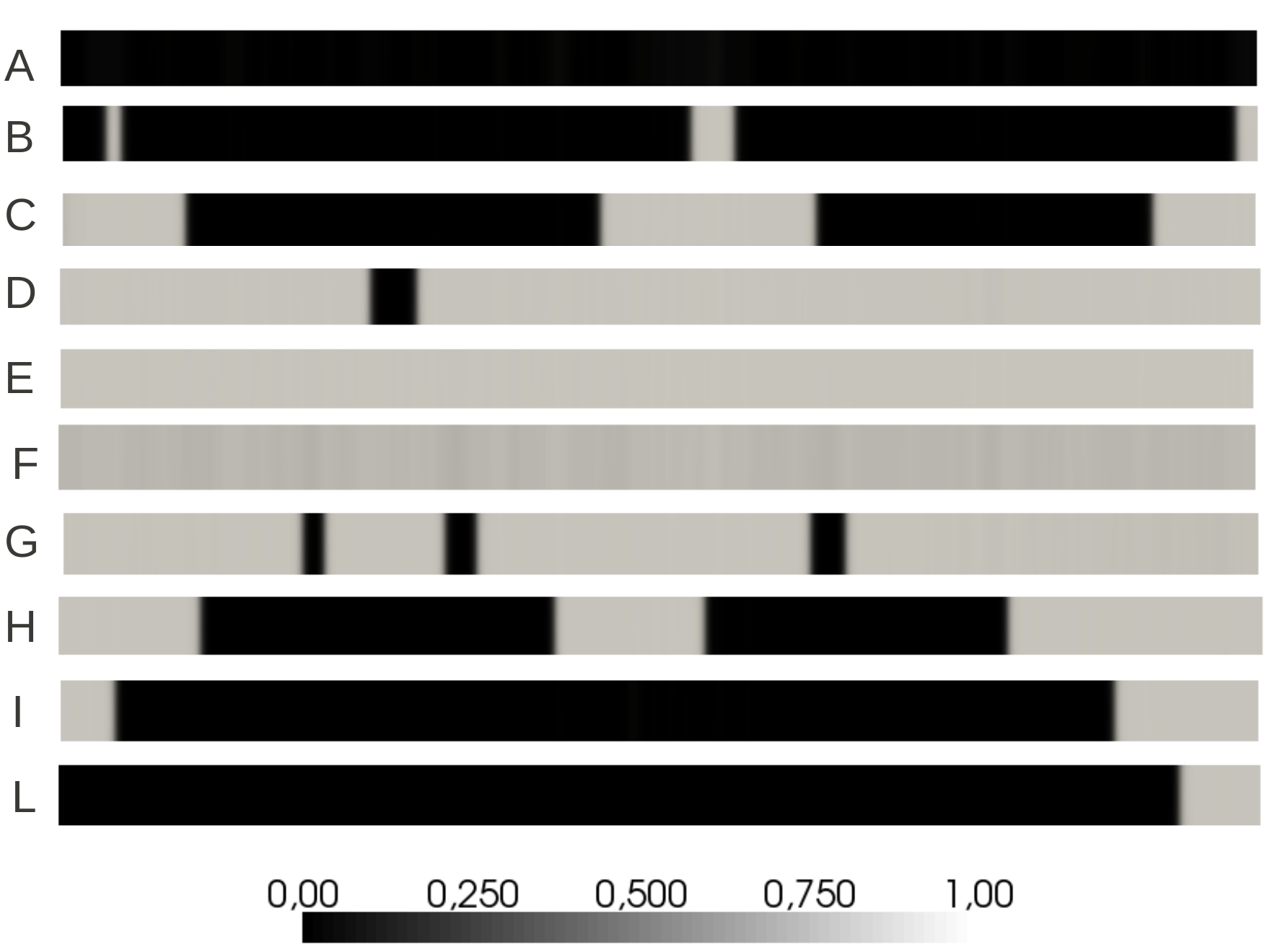}
\end{tabular}
\end{center}
\caption{Tensile test at nominal strain rate
          $\dot{\epsilon} = 3.3 \times 10^{-4} \text{ s}^{-1}$, initial temperature
          $\theta_0 = 296 \text{ K}$ and heat transfer
          coefficient $h = 10 \text{ W/(m}^{2} \text{ K)}$: (a) nominal
          stress-strain diagram; (b) phase morphology at different stages.}
\label{fig:baseline}
\end{figure}

The nominal stress-strain diagram resulting form the simulation of a tensile
test is reported in Fig. \ref{fig:baseline}(a), with the related microstructure
evolution depicted in Fig. \ref{fig:baseline}(b). Simulations are performed at
a nominal strain-rate $\dot{\epsilon}=3.3\times10^{-4} \text{ s}^{-1}$ and with
a value of the heat transfer coefficient $h= 10 \text{ W/(m}^{2} \text{ K)}$ at
an initial temperature $\theta_0 = 296 \text{ K}$.

The specimen is initially in an unstressed configuration and in the austenitic
phase. When the
test starts, the specimen is stretched and austenite deforms elastically; the
deformation of the domain is therefore homogeneous. This behaviour is maintained
until the domain reaches the point $A$ in Fig. \ref{fig:baseline}, which is the
configuration at which austenite becomes unstable and martensite starts nucleating.
At this strain rate, three martensite spots arise; the nucleation stage is
completed at point $B$ in Fig. \ref{fig:baseline}, when three martensite bands
are clearly visible. According to Chang et al. \cite{chang:2006}, we remark
that within this context the term \textit{nucleation} refers to a macroscopic
process, i.e. the origination of the transformation fronts associated with the
macroscopic localized transformation, rather than nucleation at microscopic scale
as it is properly meant in material science. The main results of the nucleation
process are a localised deformation of the martensitic bands which causes the
stress to drop by $20 \text{ MPa}$ and the formation of a number of transition
fronts.

After originating, the fronts start propagating along the domain; this stage
corresponds to the plateau of the stress-strain curve.
Interestingly, a plot of the bulk transitional free energy function for the
stress and temperature occurring at the transition fronts during phase propagation
reveals that the transformation, once it has started, is able to proceed even if
there is an energy barrier between the local minimum and the global minimum.
Owing to the presence of the gradient term in the free energy, the points of the
domain in the local minimum (A phase during loading, M$_+$ phase during unloading),
can overcome the energy barrier and evolve towards the configuration of absolute
minimum (M$_+$ phase during loading, A phase during unloading).
The progress of the transition in the plateau regime is regulated by the interplay
between thermal and mechanical phenomena. The heat released during the phase change
at the transition front leads to an increase in temperature; consequently, the
energy barrier increases and the transition is inhibited. On the other hand,
an increase in stress causes the energy barrier to decrease and the transition
is favoured. The interaction between these two opposite
trends sets the slope of the plateau in the stress-strain diagram, which may change
considerably under different strain-rates or in different environmental conditions,
as will be shown in Sections \ref{sect:strain-rate} and \ref{sect:environment}.
During the phase propagation, front-merging phenomena may occasionally occur when
two transition fronts merge together (points $D$ and $E$ in Fig. \ref{fig:baseline})
or when one front reaches the end of the specimen (point $C$ in Fig. \ref{fig:baseline},
the left transition front reaching the left end of the domain). These events are
reflected in a sudden drop and rise of the nominal stress.

At point $E$ in Fig. \ref{fig:baseline} the specimen is fully transformed into
one variant of martensite and retrieves an elastic behaviour (from $E$ to $F$) in
which the deformation of the domain is homogeneous, corresponding to an elastic
martensite deformation.

The $M_+ \rightarrow A$ transition starts at point $F$ in Fig. \ref{fig:baseline};
due to the choice of the free energy function and of the relaxation parameter of
the TDGL equation (see Section \ref{sect:GL-pot}), it has analogous features to the
$A \rightarrow M_+$ phase transition (note that the stress rise at the beginning
of the unloading plateau is of the same magnitude of the corresponding stress drop
in the loading stage and the number of product-phase nuclei is the same). The test
ends with a residual deformation due to some residual martensite in the specimen.

\begin{figure}[h!]
\begin{center}
\includegraphics[scale=0.4]{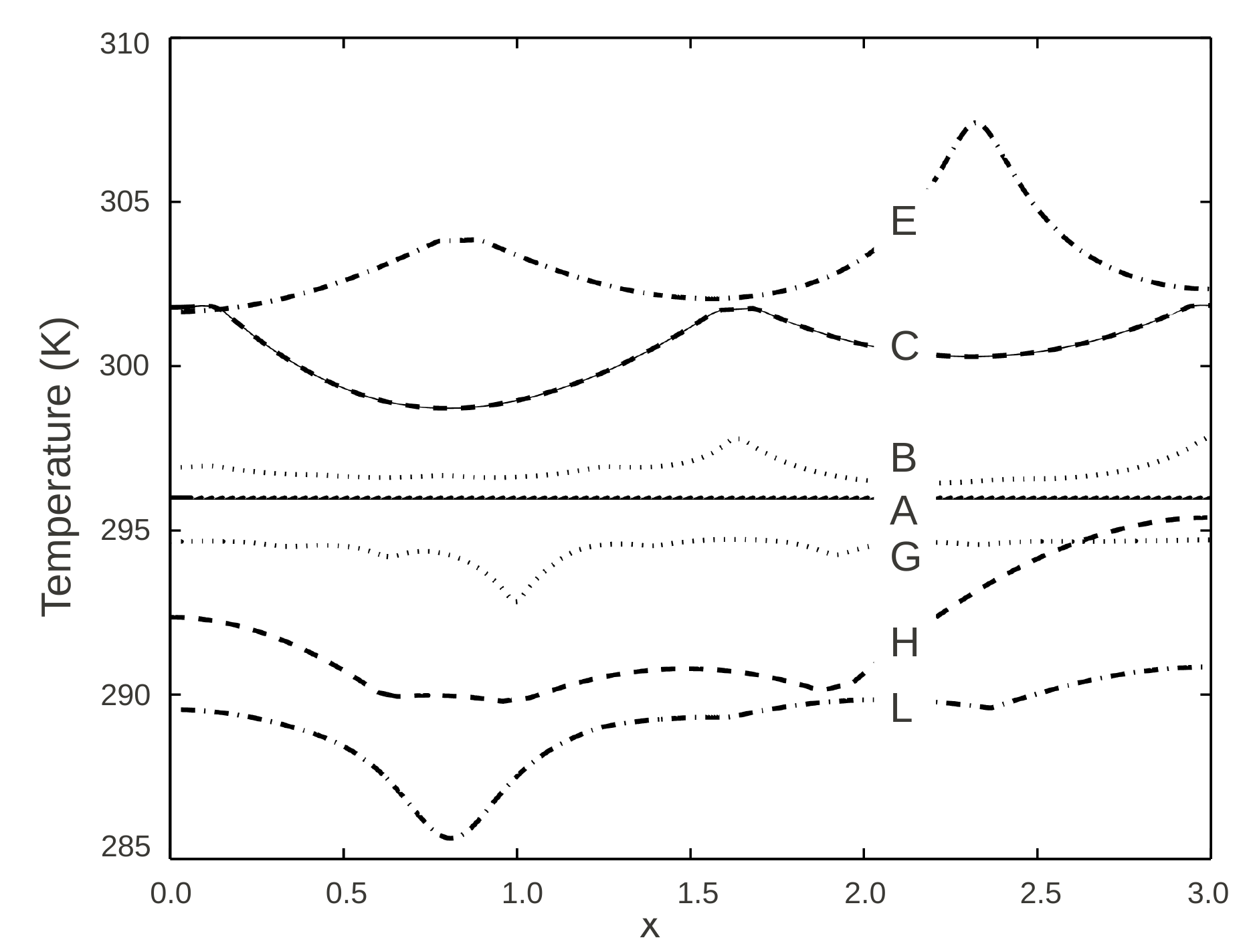}
\end{center}
\caption{Tensile test at nominal strain rate
          $\dot{\epsilon} = 3.3 \times 10^{-4} \text{ s}^{-1}$, initial temperature
          $\theta_0 = 296 \text{ K}$ and heat transfer
          coefficient $h = 10 \text{ W/(m}^{2} \text{ K)}$: temperature distribution
          along the specimen (letters refer to Fig. \ref{fig:baseline}).}
\label{fig:baseline_temp}
\end{figure}

The $A \rightarrow M_+$ transition is exothermic, hence the specimen experiences
self-heating in the neighbourhood of the transition fronts due to both latent heat
release and dissipation; furthermore, there is an overall increase in temperature
due to thermal conduction within the domain.
Opposite considerations apply to the case of the $M_+ \rightarrow A$ transformation.
The temperature distribution along the specimen is shown in Fig.
\ref{fig:baseline_temp} for some of the points marked on the stress-strain diagram
in Fig. \ref{fig:baseline}. The curve marked with $A$ refers to the onset of
the transition; until this event, the domain has deformed elastically, hence
there is no increase in temperature with respect to the initial condition.
After the nucleation, three temperature peaks rise, as depicted in curve $C$; the
temperature profile exhibits then as many fronts as the phase transition fronts,
because of the heat released during the phase change. When two propagating fronts
merge, the released heat becomes more significant and the local increase in
temperature is more pronounced; this is the situation depicted in curve $E$ in
Fig. \ref{fig:baseline_temp}, where the maximum temperature of the test is
experienced. The values are quantitatively in good agreement with the experimental
observations \cite{he:2010_2}.
The opposite behaviour is observed in the $M_+ \rightarrow A$ transformation
(curves $G$, $H$ and $L$ in Fig. \ref{fig:baseline_temp}) during which the
temperature decreases even below the initial temperature as a consequence of the
latent heat absorbed in the transition from martensite to austenite.
%

%-------------------------------------------------------------------------------
\subsection{Tensile tests at different values of the initial temperature}
\label{sect:temp-test}
\begin{figure}[h!]
\begin{center}
\includegraphics[scale=0.5]{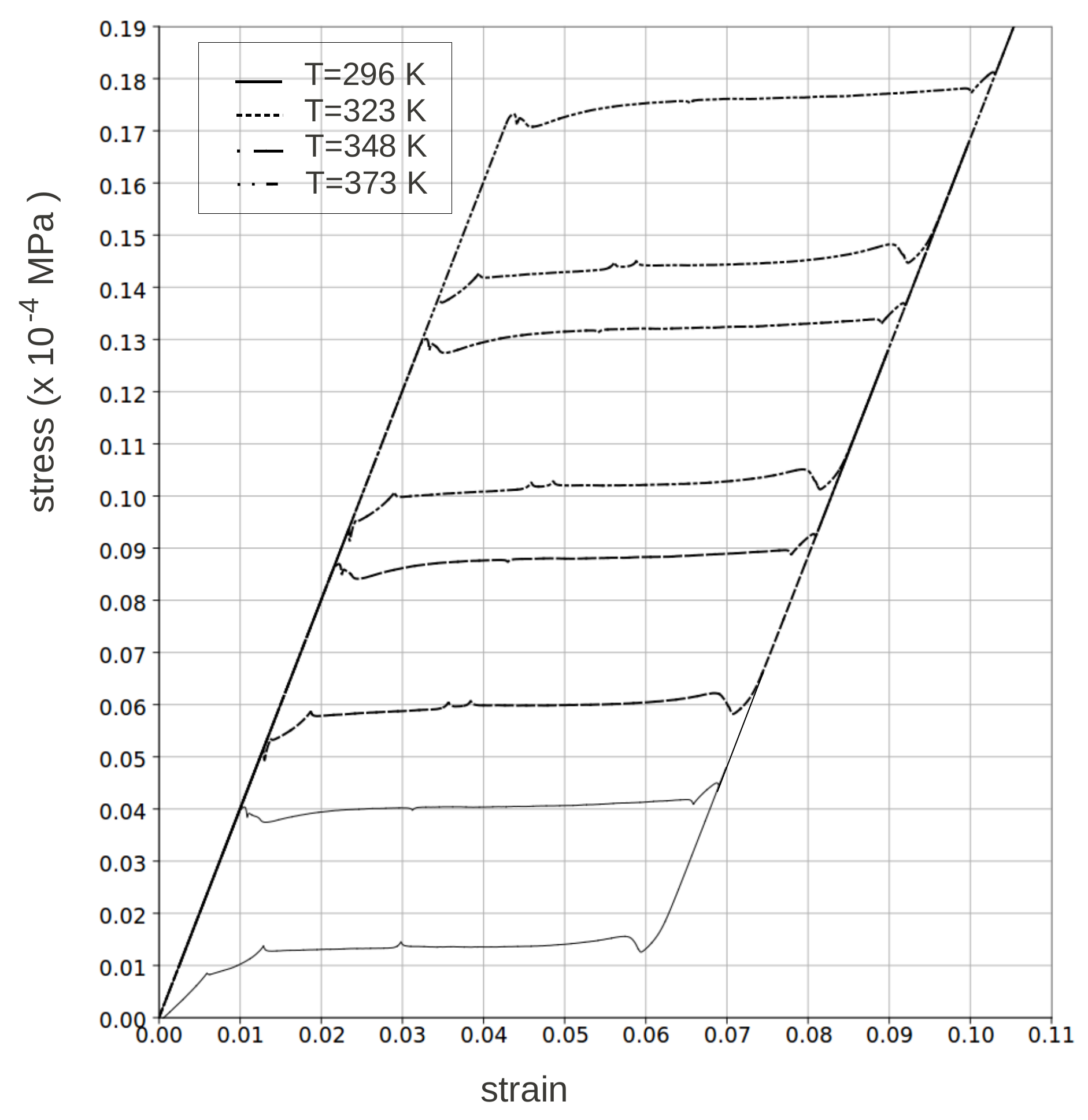}
\end{center}
\caption{Tensile tests at nominal strain rate
          $\dot{\epsilon} = 3.3 \times 10^{-4} \text{ s}^{-1}$ and heat transfer
          coefficient $h = 10 \text{ W/(m}^{2} \text{ K)}$: nominal stress-strain
          diagram for different values of the initial temperature.}
\label{fig:stress-strain-temp}
\end{figure}

The nominal stress-strain diagrams depicted in Fig. \ref{fig:stress-strain-temp}
show the results of tensile tests simulated with different values of the initial
temperature, which ranged from $\theta_0 = 296 \text{ K}$ to $\theta_0 = 373 \text{ K}$.

According to the phase diagram in Fig. \ref{fig:phase-d}, a higher value of the
temperature implies a higher stress required for the macroscopic domain nucleation.
As a result, Fig. \ref{fig:stress-strain-temp} shows that as the initial temperature
is higher, the hysteresis cycle appears shifted towards higher values of stress
and the elastic behaviour is retrieved after the hysteresis loop is completed.
Note that the area of the hysteresis loops, as well as the number of product-phase
nuclei arising at the onset of the transition are independent of the value of the
initial temperature.
%

%-------------------------------------------------------------------------------
\subsection{Tensile tests at different values of the maximum deformation}
\label{sect:max-def}
\begin{figure}[h!]
\begin{center}
\begin{tabular}{cc}
  (a)&\includegraphics[scale=0.5]{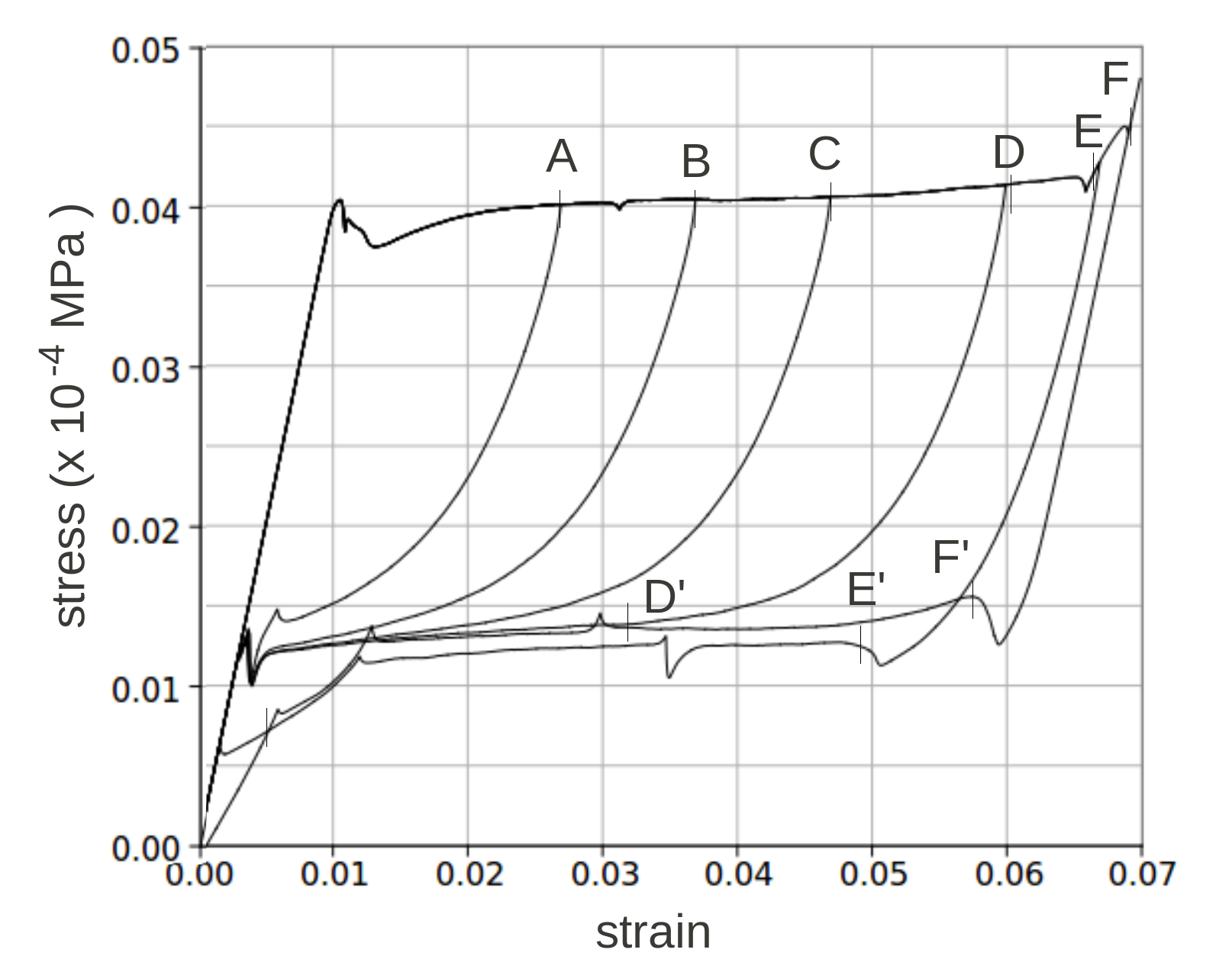} \\
  (b)&\includegraphics[scale=0.4]{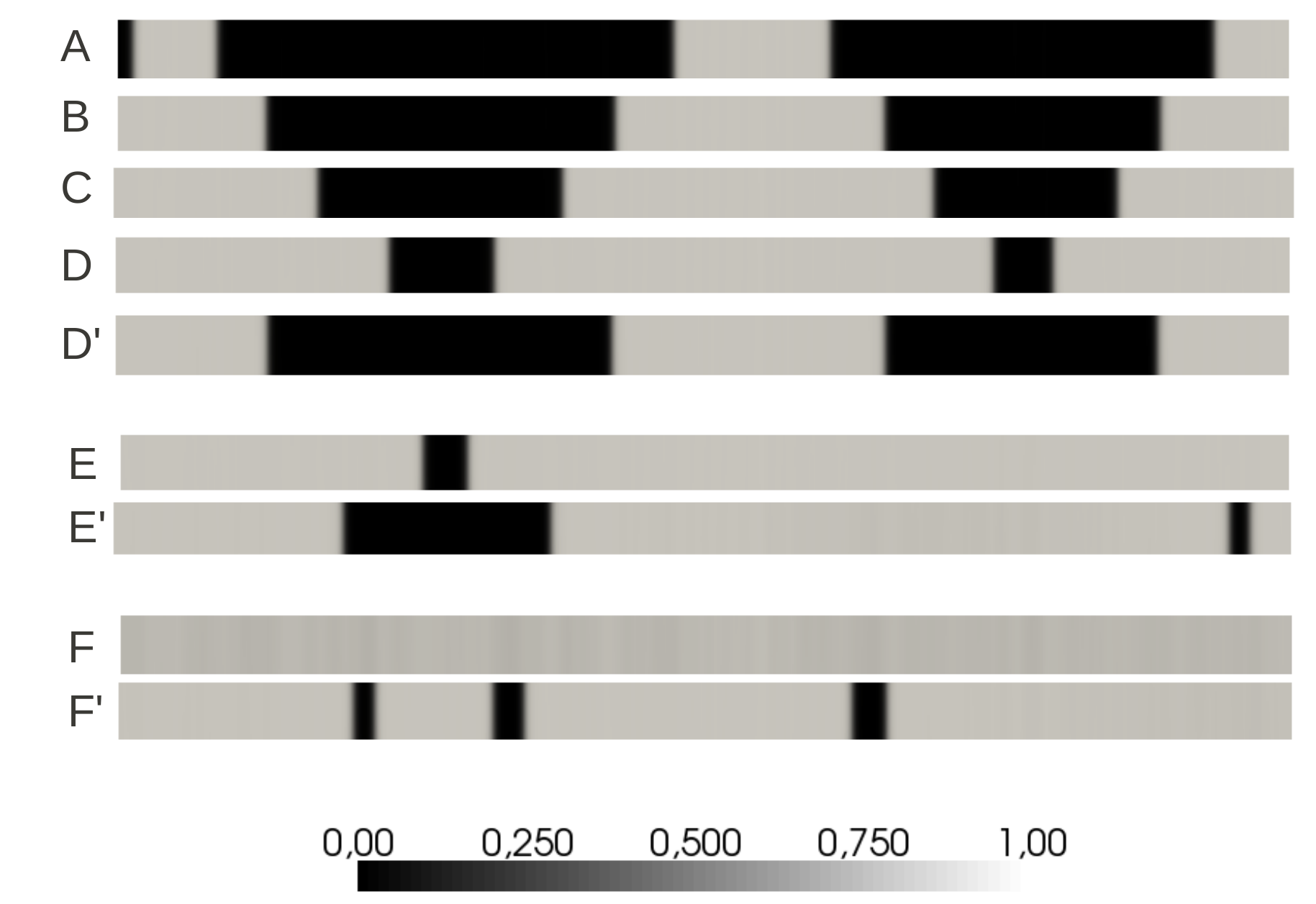}
\end{tabular}
\end{center}
\caption{Tensile tests at nominal strain rate
          $\dot{\epsilon} = 3.3 \times 10^{-4} \text{ s}^{-1}$, initial temperature
          $\theta_0 = 296 \text{ K}$ and heat transfer
          coefficient $h = 10 \text{ W/(m}^{2} \text{ K)}$ for different values
          of the maximum deformation: (a) nominal stress-strain diagram; (b) phase
          morphology at different stages.}
\label{fig:interrupted}
\end{figure}

To evaluate the behaviour of the specimen under partial-loading conditions, tensile
tests are performed at different values of the maximum imposed nominal strain,
which ranged from $\epsilon = 0.07$ to $\epsilon = 0.027$.
Fig. \ref{fig:interrupted} shows the resulting stress-strain diagrams and the
morphology of the specimen at interesting stages of the tests.

It can be observed that if the unloading begins before any domain merging, the
reverse transformation consists in a propagation of the already formed transition
fronts, rather than involving a re-nucleation process (cases $A$, $B$, $C$ and
$D$ of Fig. \ref{fig:interrupted}).
If the loading is otherwise interrupted after there has been a merging of two
fronts, there can be re-nucleation. Considering case $E$ of Fig.
\ref{fig:interrupted}, at the end of the loading stage there is only one austenite
domain left; the reverse transition starts therefore with the nucleation of one
austenite domain ($E'$) and the transformation proceeds with two propagating
domains.

Case $F$, instead, depicts the situation in which the specimen is fully transformed
into martensite upon loading; as a consequence, the reverse transition must start
with the nucleation of austenite domains. As can be seen in Fig.
\ref{fig:interrupted}(b), in this case three nuclei originate and consequently the
rise in stress due to the nucleation is higher than in case $E$.

In much of the data available in literature regarding SMA, the specimen is fully
transformed into martensite only in the gauge section; two austenitic domains
remain at the two ends of the specimen, and the reverse transformation starts
with the propagation of the two existing fronts without re-nucleation \cite{zhang:2010}.
The mechanical response of the specimen suffers therefore from an experimental
artifact and the onset of the reverse transformation does not exhibit any stress
rise due to nucleation. This condition is replicated in cases form $A$ to $D$ of
Fig. \ref{fig:interrupted}.
To overcome this issue and ensure product-phase nucleation in both loading and
unloading stages, Iadicola and Shaw \cite{iadicola:2007} make use of a thermoelectric
device to adjust the temperature at the clamped ends of the specimen in order to
ensure the transition to start within the gauge length and obtain a stress-strain
response similar to that described in case $F$ of Fig. \ref{fig:interrupted}.

%-------------------------------------------------------------------------------
\subsection{Influence of the strain-rate on the behaviour of the model}
\label{sect:strain-rate}
To explore the ability of the proposed model to reproduce the rate-dependent
behaviour experimentally observed in SMA \cite{zhang:2010}, tensile tests at
different values of the nominal imposed strain rate were performed; the value of
the strain rate ranged form $\dot\epsilon = 3.3 \times 10^{-4} \text{ s}^{-1}$ to
$\dot\epsilon = 3.3 \times 10^{-2} \text{ s}^{-1}$, while the initial temperature
and the maximum imposed displacement were the same for all the test-cases
($\theta_0 = 296 \text{ K}$, $\epsilon = 0.07$).
\begin{figure}[h!]
\begin{center}
\begin{tabular}{cc}
  (a)&\includegraphics[scale=0.5]{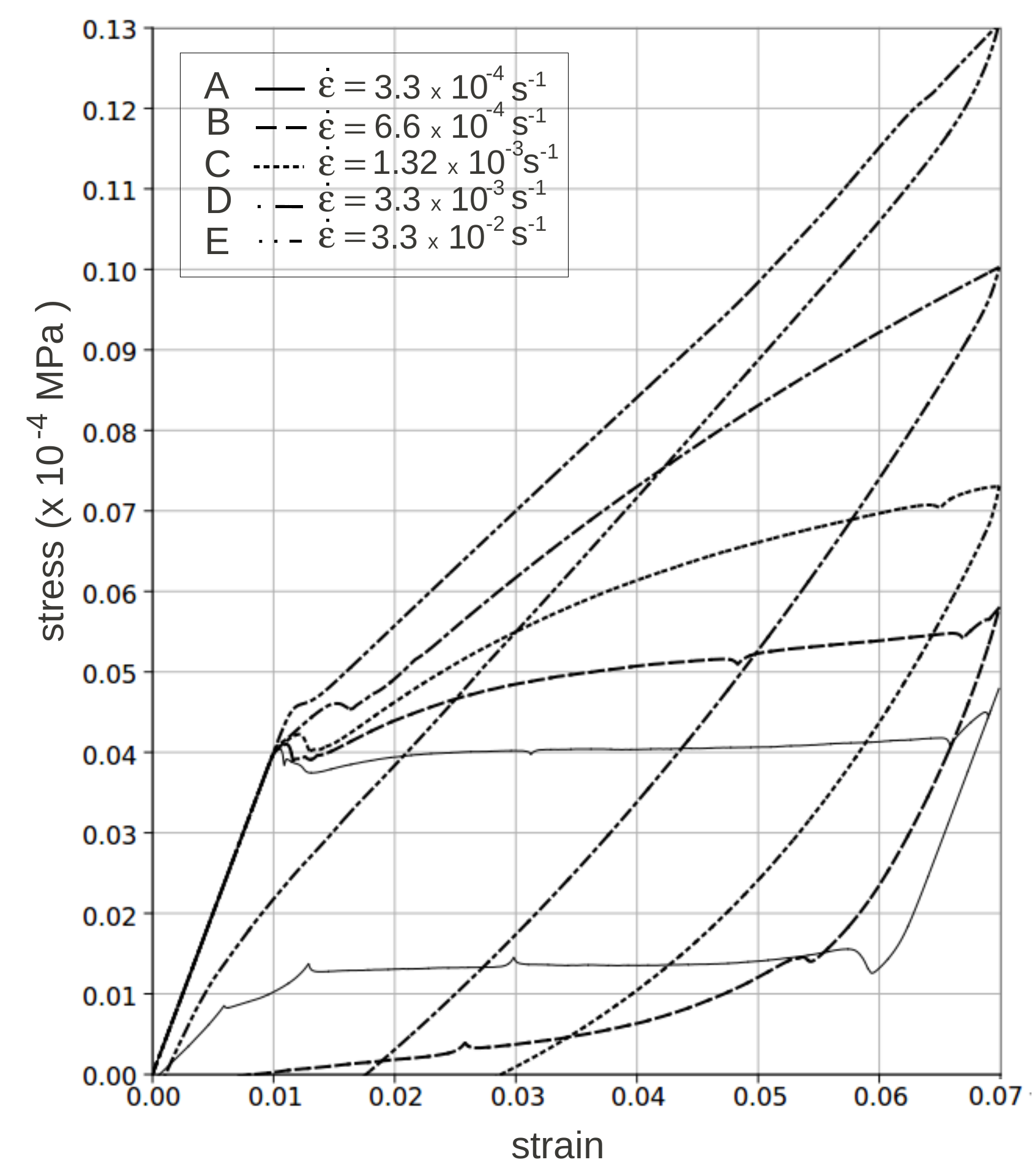} \\
  (b)&\includegraphics[scale=0.4]{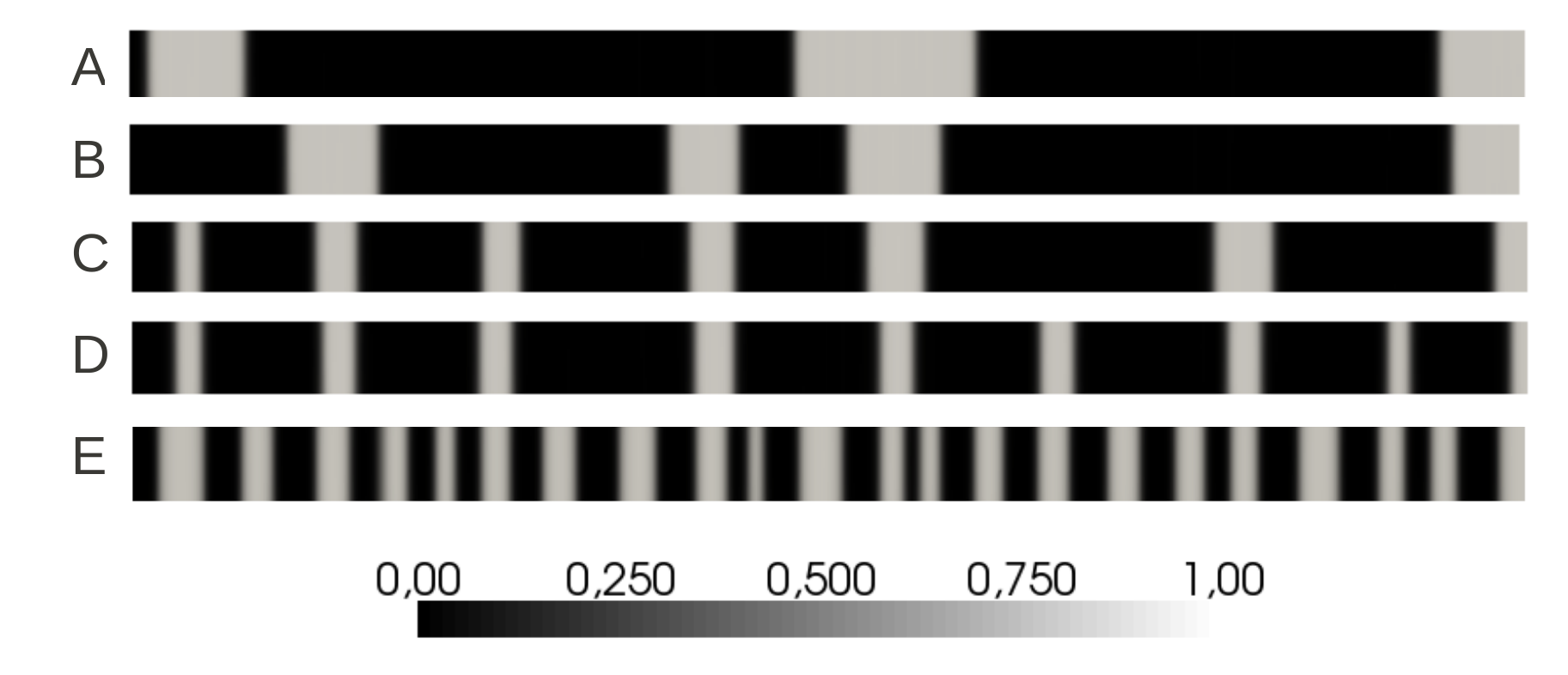}
\end{tabular}
\end{center}
\caption{Tensile tests at different values of nominal strain rate, initial temperature
          $\theta_0 = 296 \text{ K}$ and heat transfer
          coefficient $h = 10 \text{ W/(m}^{2} \text{ K)}$: (a) nominal stress-strain
          diagram; (b) phase morphology after $M_+$ nucleation.}
\label{fig:stress-strain-rate}
\end{figure}

\begin{figure}
\begin{center}
\begin{tabular}{cc}
 \includegraphics[scale=0.35]{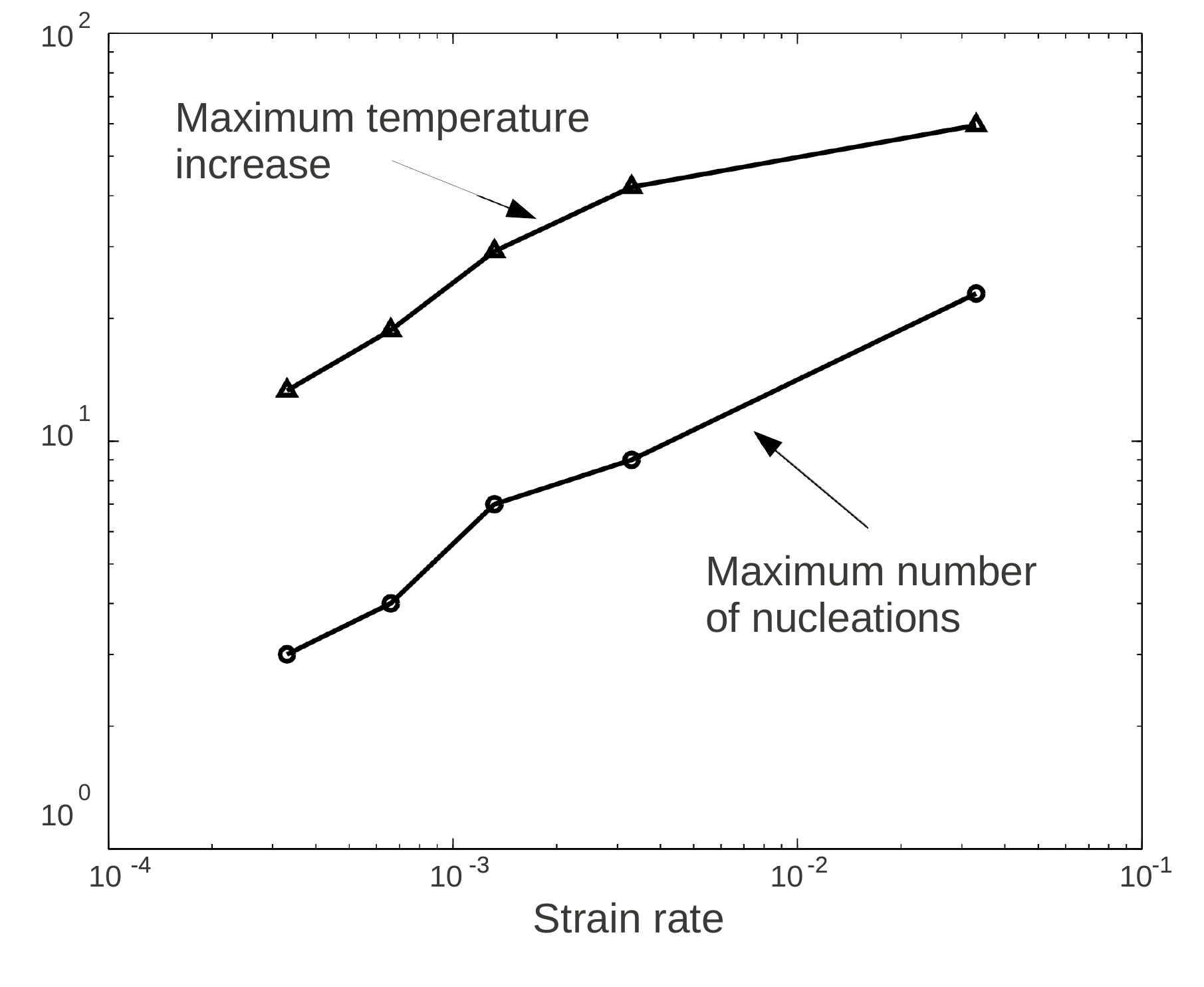} & \includegraphics[scale=0.35]{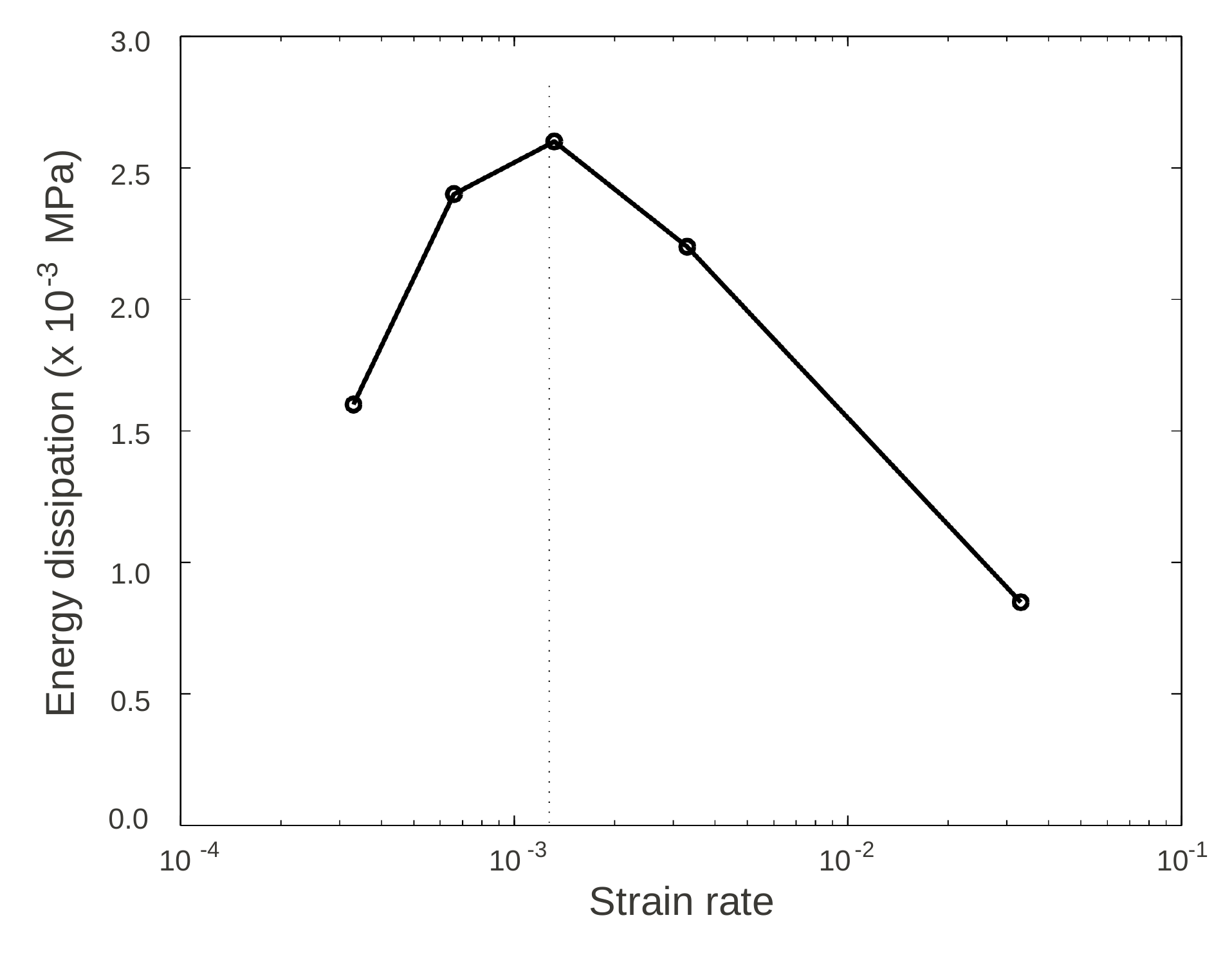} \\
  (a) & (b)
\end{tabular}
\caption{(a) maximum increase in temperature within the specimen and number of
          product-phase nuclei as a function of the test strain rate.
          (b) Energy dissipation as a function of the test strain-rate.}
\label{fig:temp-diss-rate}
\end{center}
\end{figure}

As the strain-rate becomes higher, the stress at which nucleation starts increases
(Fig. \ref{fig:stress-strain-rate}(a));
as a consequence, the driving force for the phase transition is greater and more
product-phase nuclei will originate (Fig. \ref{fig:stress-strain-rate}(b)). Fig.
\ref{fig:temp-diss-rate}(a) depicts the maximum number of product phase
nuclei as a function of the nominal imposed strain rate; the trend resulting from
the simulations is exponential, in agreement with the experimental observation of
He and Sun \cite{he:2010_2} and the numerical simulations of Iadicola and Shaw
\cite{iadicola:2004}.

Fig. \ref{fig:stress-strain-rate}(a) also shows that the discontinuities in the
stress-strain curve induced by nucleation and front merging tend to vanish at high
strain rates, because the loading time scale gets closer to the time scale of the
phase transition phenomena.

One of the main effects of the strain-rate on the mechanical behaviour of the specimen
is the change in the slope of the stress plateau during the loading stage, which
results from an interplay between transitional, thermal and mechanical phenomena.
As the strain rate becomes higher, the heat released due to the phase transition
increases, but there is less time for such heat to be transferred to the exterior,
because the duration of the loading stage is shorter; as a consequence, at a given
strain the temperature increases.
In order for the transition to proceed, the stress must therefore increase, which
makes the slope of the stress plateau increase.
Moreover, if the stress at the end of the loading stage is higher, the elastic
deformation of the specimen is greater, hence less deformation from the phase
transition is required to achieve the maximum imposed deformation and more
untransformed austenite remains.
Furthermore, upon unloading the stress rise described in Section \ref{sect:baseline}
is absent at higher stress rate; this effect is due to the fact that at the fixed
maximum strain there remains more untransformed austenite, hence no re-nucleation
occurs.

As a direct evidence of the scarcity of time for the heat generated within the
specimen to be transferred outside it, the maximum temperature reached at the
end of the loading stage is higher for higher strain rates.
The values of the maximum temperature increase within the specimen are reported in
Fig. \ref{fig:temp-diss-rate}, in which it can be noted that higher
strain-rates correspond to greater temperature increases.

The strain-rate has a strong impact also on the damping capacity of the
specimen; a measure of this property is the area of the hysteresis cycle in the
stress-strain diagram, which represents the mechanical energy dissipated during
the test. Fig. \ref{fig:temp-diss-rate}(b) shows that the dissipated
energy changes non-monotonically with the strain rate.
As observed by Zhang et al. \cite{zhang:2010}, the non-monotone trend of the
dissipated energy is due to the fact that during the $A \rightarrow M_+$
transformation the slope of the stress plateau increases monotonically with the
strain rate, hence the transition occurs for higher values of the stress; on the
other hand, the $M_+ \rightarrow A$ transition has a non-monotone trend \cite{zhang:2010}.

%-------------------------------------------------------------------------------
\subsection{Influence of the heat transfer conditions on the behaviour of the model}
\label{sect:environment}
The rate-dependent behaviour of SMA stems from an interaction between transitional,
thermal and mechanical phenomena; hence, changing the conditions under which the
generated heat is exchanged with the exterior affects the mechanical response of
the material \cite{he:2010_1}.
To highlight this point, tensile tests simulating different heat exchange conditions
were performed; this was achieved by changing the value of the heat transfer
coefficient, which ranged form $h = 2 \text{ W/(m}^{2} \text{ K)}$ to
$h = 50 \text{ W/(m}^{2} \text{ K)}$.
For all the test-cases, the initial temperature and the imposed strain rate were
the same ($\theta_0 = 296 \text{ K}$, $\dot \epsilon = 3.3 \times 10^{-3} \text{ s}^{-1}$).
\begin{figure}[h!]
\begin{center}
\includegraphics[scale=0.5]{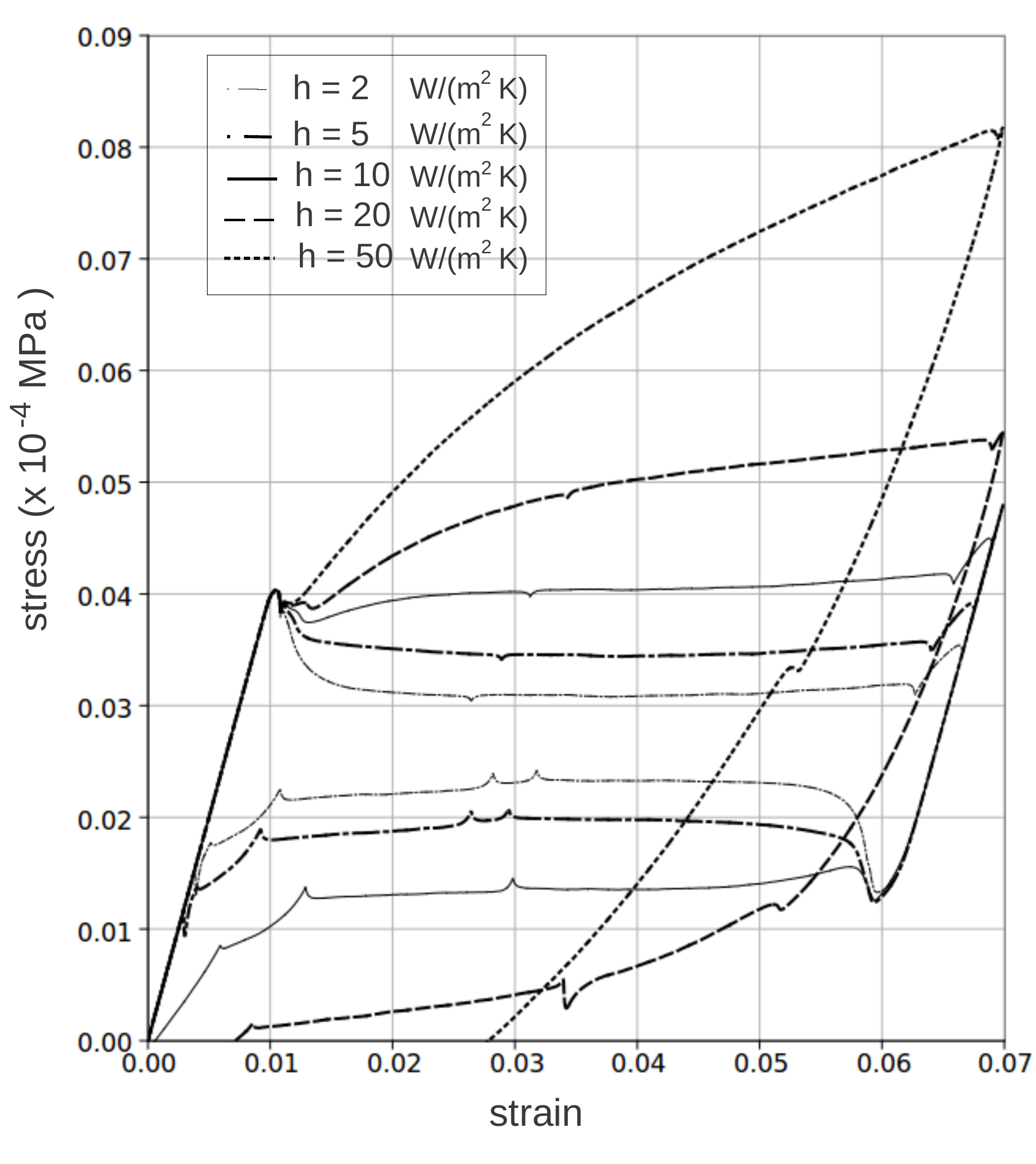}
\end{center}
\caption{Tensile tests at nominal strain rate
          $\dot{\epsilon} = 3.3 \times 10^{-3} \text{ s}^{-1}$ and initial
          temperature $\theta_0 = 296 \text{ K}$: nominal stress-strain
          diagram for different values of the heat transfer coefficient.}
\label{fig:stress-strain-heat}
\end{figure}

The nominal stress-strain diagram for the different tests is reported in Fig.
\ref{fig:stress-strain-heat}.
A low value of the heat transfer coefficient causes an increase in the time needed
for the generated heat to be transferred to the exterior. If the value of the
coefficient is very low, the test proceeds in adiabatic condition, as the time
scale of the heat transfer phenomena becomes much larger than the test time scale.
As a result, in the loading stage the temperature at the transition front increases
considerably and a higher stress is required for the transition to proceed.
On the contrary, if the heat transfer coefficient is high, the generated heat is
transferred to the exterior in short time and the test proceeds nearly isothermally;
under this condition, the temperature does not contribute to increase the energy
barrier and the transition can therefore proceed at a low value of the stress.
As a result, the stress plateau is horizontal and the difference between the
nucleation stress and the propagation stress is relevant.

%-------------------------------------------------------------------------------
\subsection{Concurrent effect of ambient condition and strain-rate on the behaviour of the model}
\label{sect:env-rate}
So far, the effects of the strain-rate and of the heat transfer on the
mechanical behaviour of SMA have been investigated disjointly; in this section,
we discuss the concurrent influence of these two aspects on the damping capacity
and on the maximum temperature experienced during the tests.

\begin{figure}
\begin{center}
\begin{tabular}{cc}
 \includegraphics[scale=0.35]{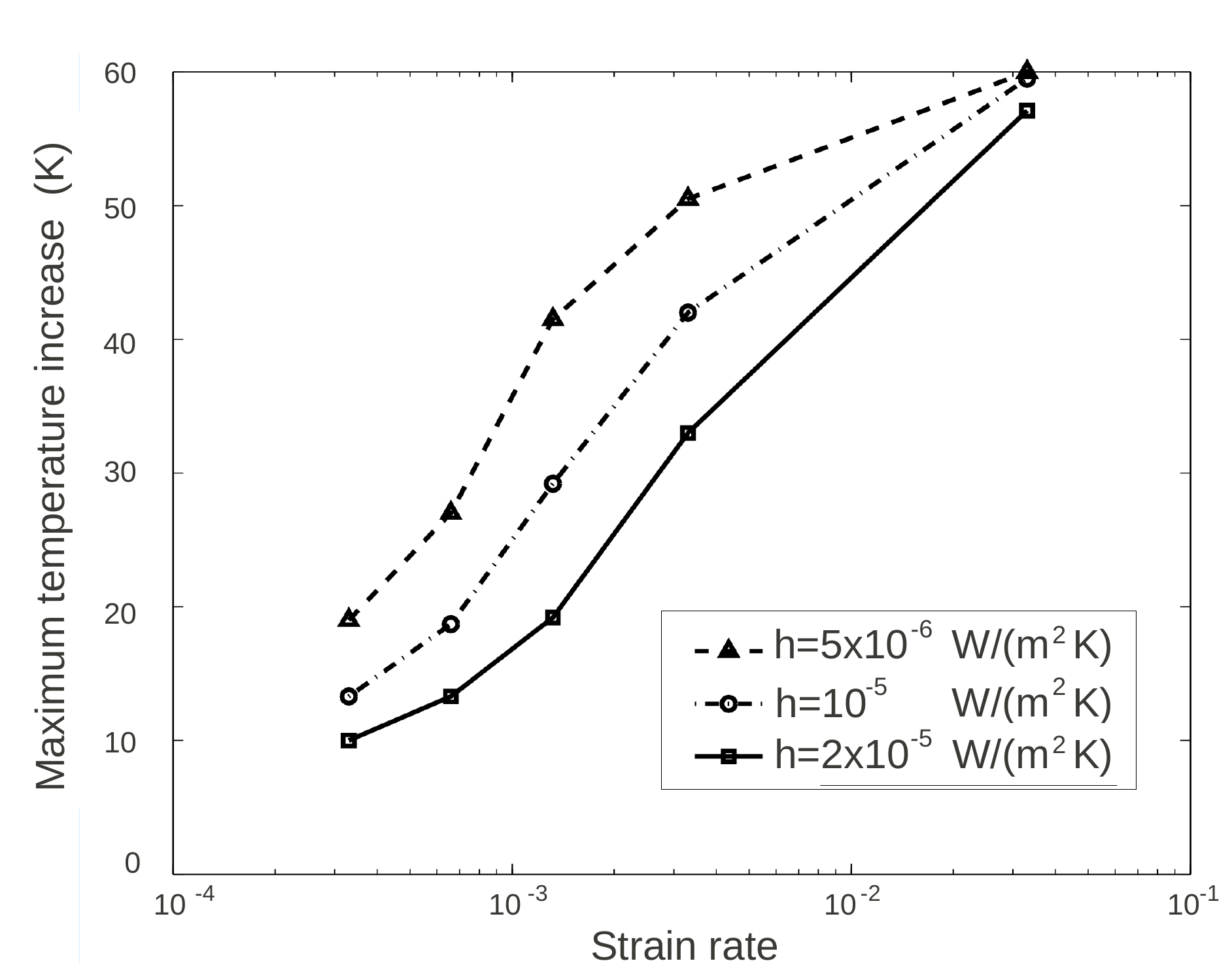} & \includegraphics[scale=0.35]{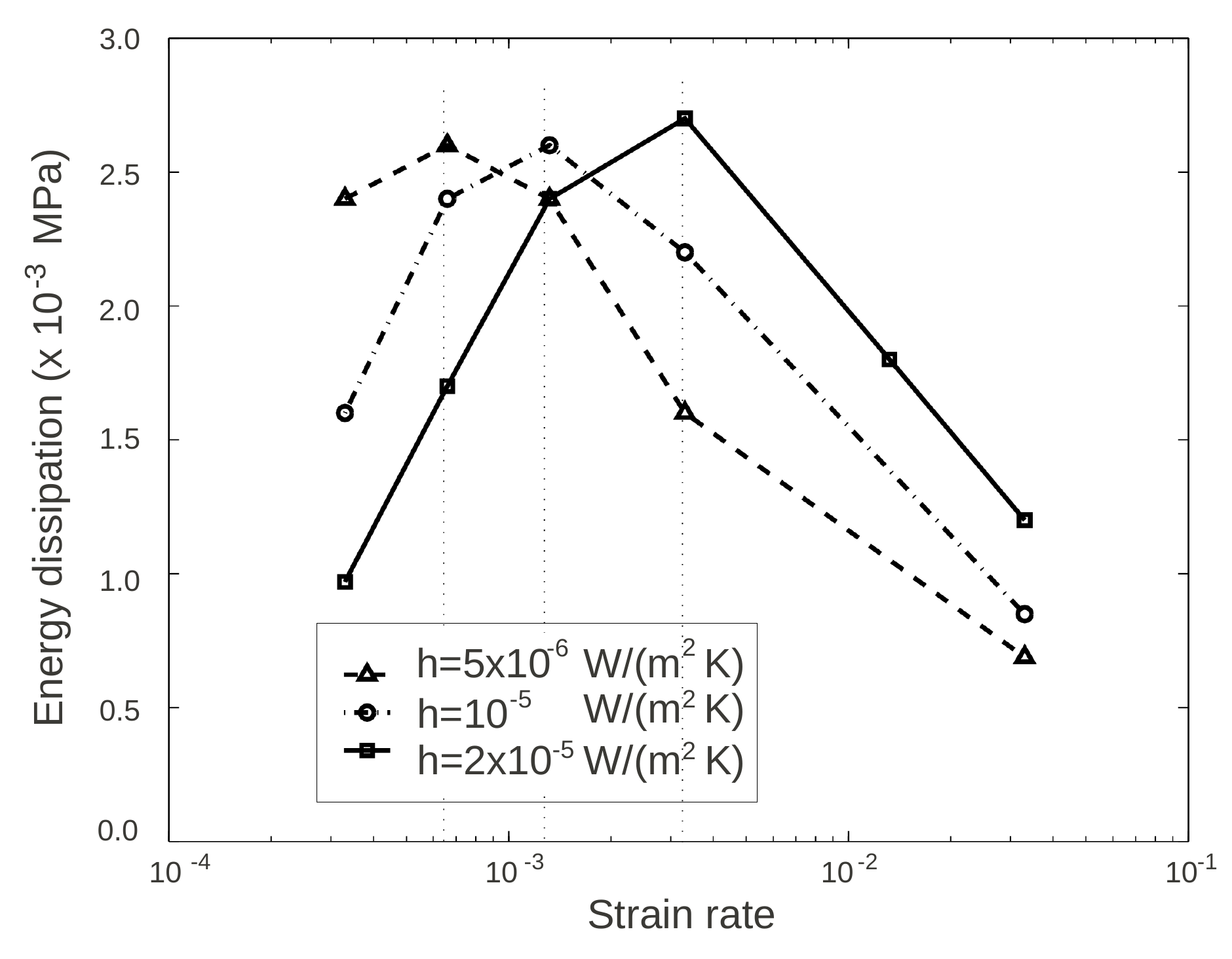} \\
  (a) & (b)
\end{tabular}
\caption{(a) maximum increase in temperature within the specimen and (b) energy
          dissipation as a function of the test strain-rate for different values
          of the heat transfer coefficient.}
\label{fig:temp-diss-rate-heat}
\end{center}
\end{figure}

Fig. \ref{fig:temp-diss-rate-heat}(a) depicts the maximum increase in temperature
during the simulated tensile test for different values of the heat transfer
coefficient and of the imposed strain rate. Both a high value of the strain rate
and a low value of the heat transfer coefficient cause the maximum temperature
inside the specimen to increase, as a smaller amount of the generated heat can be
transferred to the exterior. Note also that at high strain rates the temperature
increase within the specimen tends to become insensitive to the heat transfer
coefficient, as the adiabatic condition is approached regardless heat transfer
phenomena.

The non-monotone behaviour of the damping capacity for different values of the
strain rate was analysed in Section \ref{sect:strain-rate}; the same test was performed
for different values of the heat transfer coefficient ($h = 5 \text{ W/(m}^{2} \text{ K)}$,
$h = 20 \text{ W/(m}^{2} \text{ K)}$) and the results obtained are shown in Fig.
\ref{fig:temp-diss-rate-heat}(b). It can be observed that a high value of the heat exchange
coefficient causes the peak of the dissipated energy to be located at higher values
of the strain rate, owing to the fact that heat transfer towards the exterior is
favoured and the adiabatic condition is shifted towards higher values of the
strain rate, thus. Note that, for the range of values of interest in this study,
there is no significant effect of the heat transfer coefficient on the maximum
value of the dissipated energy.
\begin{figure}[h!]
\begin{center}
\begin{tabular}{cc}
  (a) & \includegraphics[scale=0.5]{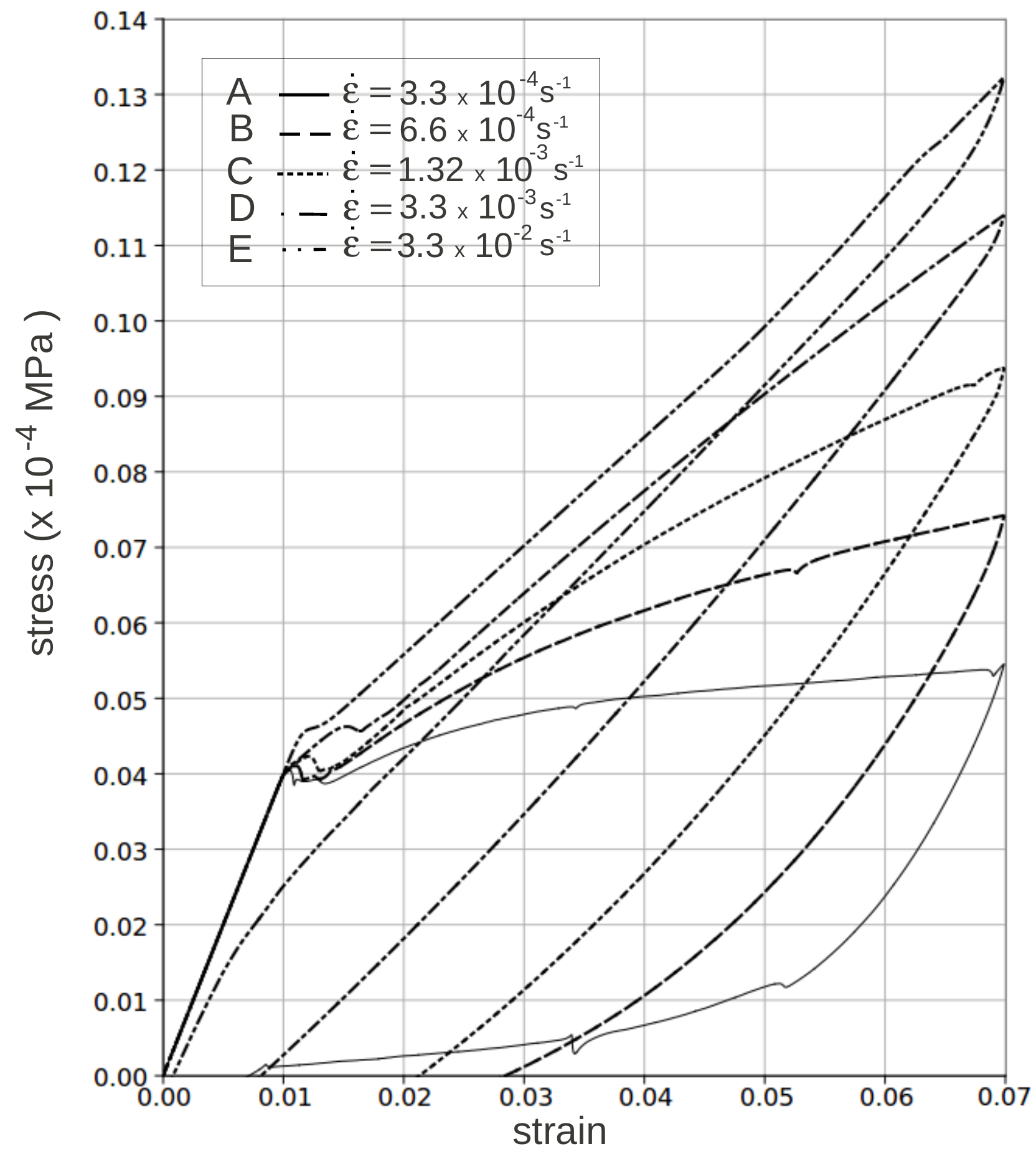} \\
  (b) & \includegraphics[scale=0.4]{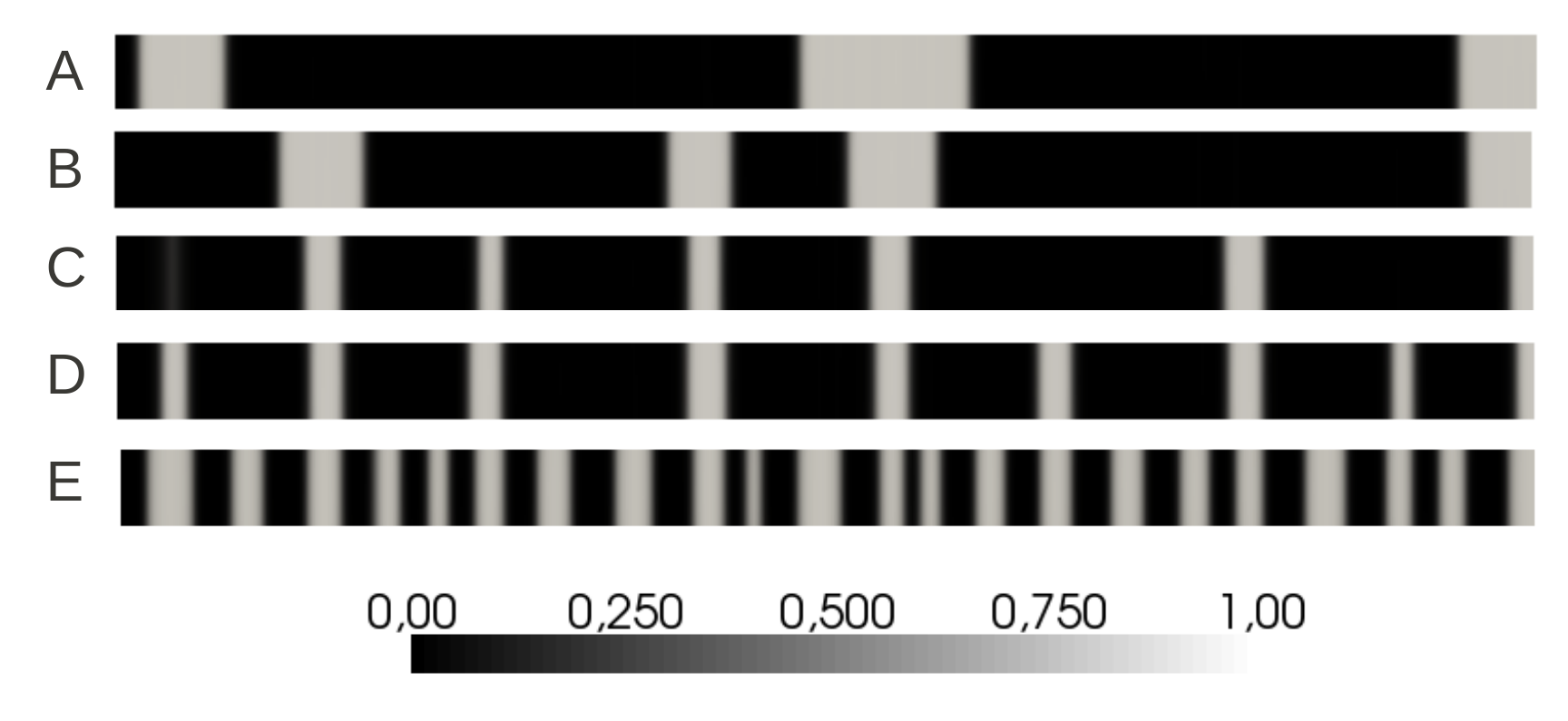}
\end{tabular}
\end{center}
\caption{Tensile tests at different values of nominal strain rate, initial temperature
          $\theta_0 = 296 \text{ K}$ and heat transfer
          coefficient $h = 5 \text{ W/(m}^{2} \text{ K)}$: (a) nominal stress-strain
          diagram; (b) phase morphology after $M_+$ nucleation.}
\label{fig:stress-strain-rate-h5}
\end{figure}
\begin{figure}[h!]
\begin{center}
\begin{tabular}{cc}
  (a) & \includegraphics[scale=0.5]{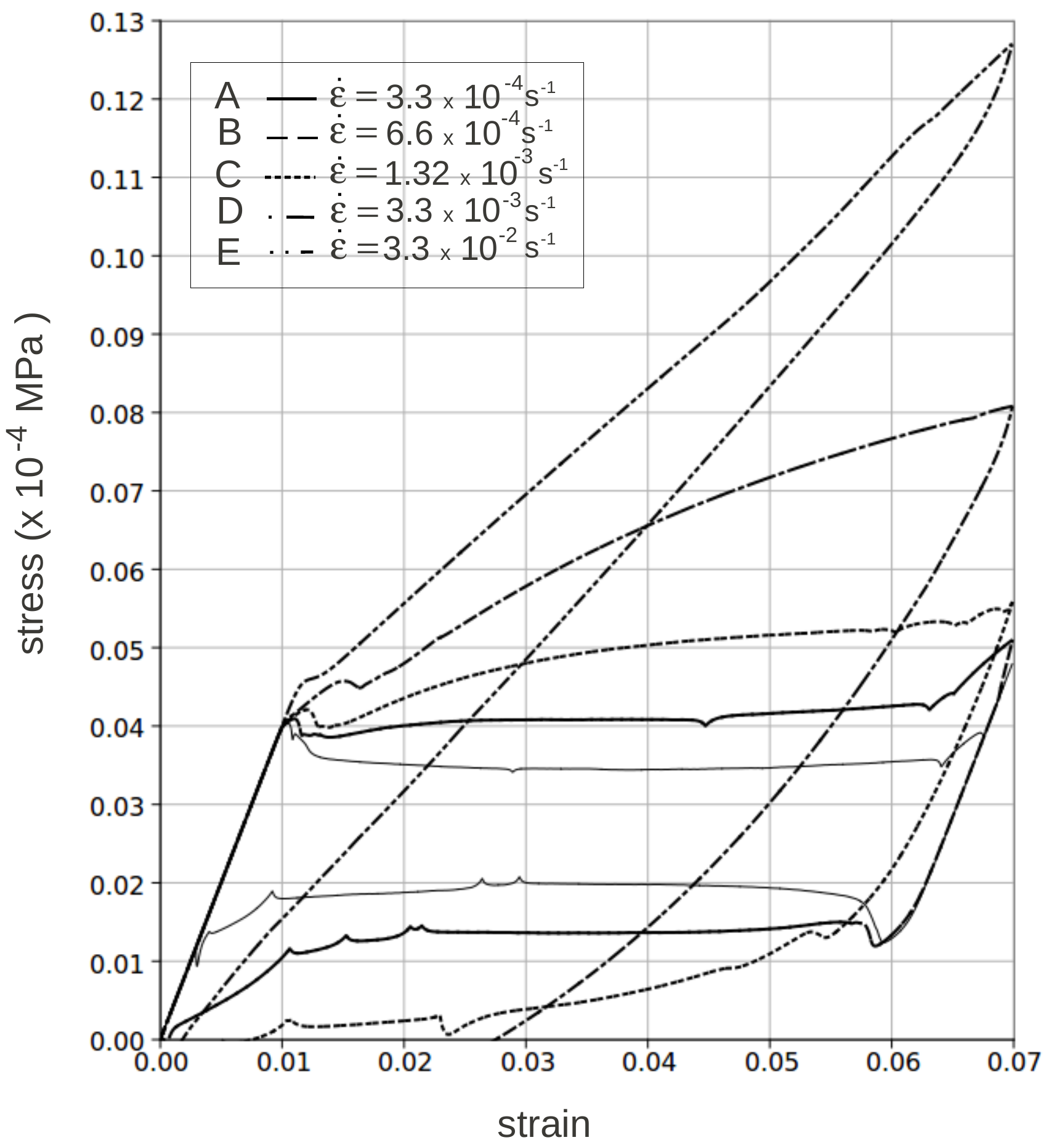} \\
  (b) & \includegraphics[scale=0.4]{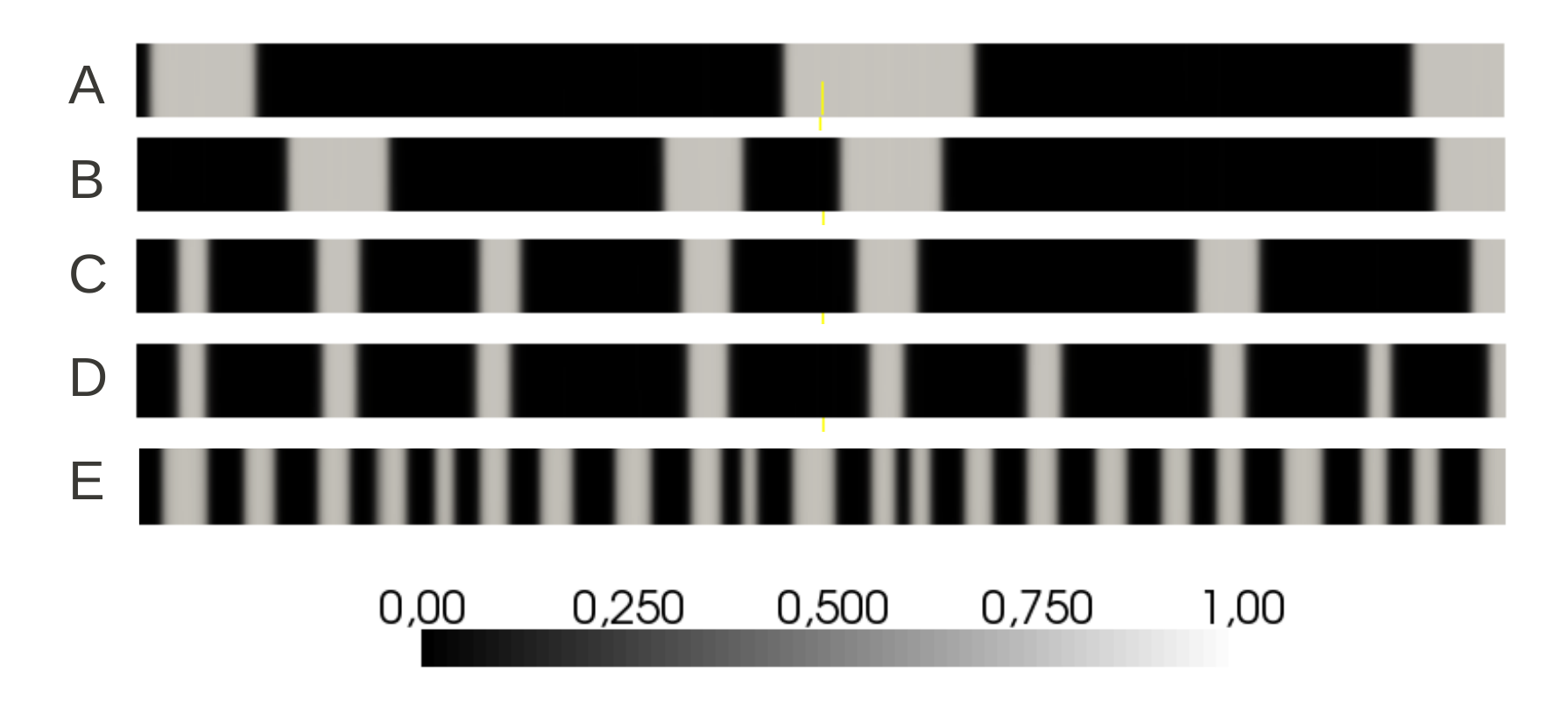}
\end{tabular}
\end{center}
\caption{Tensile tests at different values of nominal strain rate, initial temperature
          $\theta_0 = 296 \text{ K}$ and heat transfer
          coefficient $h = 20 \text{ W/(m}^{2} \text{ K)}$: (a) nominal stress-strain
          diagram; (b) phase morphology after $M_+$ nucleation.}
\label{fig:stress-strain-rate-h20}
\end{figure}

The stress-strain diagrams referring to the tests performed at different strain
rates for the values of the heat transfer coefficient $h = 5 \text{ W/(m}^{2} \text{ K)}$ and
$h = 20 \text{ W/(m}^{2} \text{ K)}$ are reported in Figs. \ref{fig:stress-strain-rate-h5} and
\ref{fig:stress-strain-rate-h20}, respectively, together with the maximum number of
martensite nuclei originated during the loading stage. The behaviour is the same
described in Section \ref{sect:strain-rate} for the case $h = 10 \text{ W/(m}^{2} \text{ K)}$.

%%%%%%%%%%%%%%%%%%%%%%%%%%%%%%%%%%%%%%%%%%%%%%%%%%%%%%%%%%%%%%%%%%%%%%%%%%%%%%%%
\section{Conclusions}
\label{sect:concl}
A Ginzburg-Landau model that reproduces the most relevant macroscopic features
of the behaviour of a shape memory alloy has been considered.
The model encompasses a time-dependent Ginzburg-Landau equation for the evolution
of the phase order parameter, the balance of linear momentum and the heat equation,
in order to account for the interplay between the transitional, mechanical and
thermal aspects which strongly influences the behaviour of SMA.
Starting from the theoretical model described in Berti et al. \cite{berti:2010_2},
a new free energy based on trigonometric functions has been formulated and a
suitable form of the relaxation parameter of the TDGL equation has been proposed.
The resulting properties of the model ensure the driving force for the phase
transition from austenite to martensite to be equal to the driving force for the
transition from martensite to austenite and, consequently, provide a symmetrical
behaviour of the model upon loading and unloading.

Numerical simulations have been performed to investigate the mechanical response
of the model; tensile tests on a bar specimen have been simulated under different
conditions by changing the initial temperature, the maximum imposed strain, the
nominal imposed strain-rate and the environmental conditions.
The results obtained illustrate that the model is able to reproduce the main
experimental evidences reported in literature \cite{leo:1993, shaw:1995,
zhang:2010, he:2010_2, he:2010_1}. The simulated tensile tests show the presence
of a stress drop during the phase nucleation, which may be absent in the unloading
stage if the loading is interrupted before the completion of the phase transition;
temperature profiles consistent with the observed physical phenomena have also
been described.
The effect of the strain-rate on the mechanical response of the model have been
investigated; its impact on the number of nucleating domains, the nucleation
stress, the stress relaxation during nucleation and on the slope of the stress
plateau (strain-hardening) have been recovered.
The role of the environmental conditions in the mechanical response have also
been examined, and the concurrent influence of heat transfer conditions and
strain-rate on the damping capacity and on the overall behaviour of the model
have been described.

%%%%%%%%%%%%%%%%%%%%%%%%%%%%%%%%%%%%%%%%%%%%%%%%%%%%%%%%%%%%%%%%%%%%%%%%%%%%%%%%
\section*{Acknowledgements}
The authors wish to thank Prof. M. Fabrizio and Prof. P.G. Molari for the
inspiring discussions.

%%%%%%%%%%%%%%%%%%%%%%%%%%%%%%%%%%%%%%%%%%%%%%%%%%%%%%%%%%%%%%%%%%%%%%%%%%%%%%%%

\end{document}